\documentclass[twocolumn]{aastex631}

\newcommand{\sersic}{S\'{e}rsic}
\newcommand{\boo}{Bo{\"o}tes}
\newcommand{\booI}{Bo{\"o}tes I}
\shortauthors{M. G. Walker et al.}
\usepackage{natbib}
\usepackage{graphicx}

\newcommand{\betaBootesI}{$\beta=3.40 \pm 0.55$}
\newcommand{\gammaBootesI}{$\gamma=-0.43 \pm 0.41$}

\newcommand{\gammaBootesIV}{$\gamma=1.37 \pm 0.70$}

\newcommand{\gammaCarina}{$\gamma=0.43 \pm 0.40$}

\newcommand{\betaCraterII}{$\beta=8.37 \pm 1.24$}

\newcommand{\dlogevidenceCraterII}{$\log_{10}[E_{\alpha\beta\gamma}/E_{\rm Plum}]=1.30 \pm 0.20$}

\newcommand{\gammaDraco}{$\gamma=0.20 \pm 0.24$}

\newcommand{\betaEridanusII}{$\beta=8.40 \pm 1.17$}
\newcommand{\gammaEridanusII}{$\gamma=0.05 \pm 0.49$}

\newcommand{\gaiagammaFornax}{$\gamma=0.24 \pm 0.19$} 
\newcommand{\betaFornax}{$\beta=9.68 \pm 0.33$}
\newcommand{\gammaFornax}{$\gamma=0.18 \pm 0.13$}

\newcommand{\gammaHercules}{$\gamma=1.46 \pm 0.73$}

\newcommand{\betaLeoI}{$\beta=8.01 \pm 0.97$}
\newcommand{\gammaLeoI}{$\gamma=-0.04 \pm 0.54$}

\newcommand{\betaLeoII}{$\beta=8.78 \pm 0.78$}
\newcommand{\gammaLeoII}{$\gamma=-0.12 \pm 0.36$}

\newcommand{\betaSculptor}{$\beta=4.50 \pm 0.08$}
\newcommand{\gammaSculptor}{$\gamma=-0.15 \pm 0.18$}

\newcommand{\sextansbetadecals}{$\beta=3.70 \pm 0.10$}
\newcommand{\sextansbetadelve}{$\beta=4.26 \pm 0.33$}
\newcommand{\sextansbetapsone}{$\beta=3.86 \pm 0.99$}
\newcommand{\sextansbetasdss}{$\beta=4.94 \pm 1.58$}
\newcommand{\sextansbetagaia}{$\beta=7.28 \pm 1.43$}
\newcommand{\betaSextans}{$\beta=4.26 \pm 0.33$}
\newcommand{\gammaSextans}{$\gamma=-0.37 \pm 0.38$}

\newcommand{\betaUrsaMajorII}{$\beta=3.64 \pm 1.56$}
\newcommand{\gammaUrsaMajorII}{$\gamma=1.77 \pm 0.13$}

\newcommand{\betaUrsaMinor}{$\beta=4.57 \pm 0.11$}
\newcommand{\gammaUrsaMinor}{$\gamma=0.14 \pm 0.13$}

\newcommand{\badnames}{Aquarius III, Bootes III, Cetus III, Hydra II, Pegasus III, Reticulum III, Sextans II, Virgo I, Virgo II, Virgo III}
\newcommand{\ngal}{61} 
\newcommand{\ngood}{51} 
 
\newcommand{\nabg}{11} 
 
\newcommand{\nbad}{10} 
 
\newcommand{\abgs}{Bootes I, Draco, Eridanus II, Fornax, Hercules, Leo I, Leo II, Sculptor, Sextans, Ursa Major II, Ursa Minor} 
\newcommand{\seglogabg}{$1.47 \pm 0.30$}
\newcommand{\seglinabg}{$29 \pm 309$}
\newcommand{\seglogplum}{$1.38 \pm 0.07$}
\newcommand{\seglinplum}{$24 \pm 4$}

\usepackage[normalem]{ulem}

\begin{document}

\title{Structural Diversity Among the Milky Way's Dwarf Spheroidal Satellites}

\correspondingauthor{Matthew G. Walker}
\email{mgwalker@cmu.edu}

\author[0000-0003-2496-1925]{Matthew G. Walker}
\affiliation{McWilliams Center for Cosmology and Astrophysics, Carnegie Mellon University, 5000 Forbes Ave, Pittsburgh, PA 15213, USA }

\author{Rapha\"el Errani}
\affiliation{McWilliams Center for Cosmology and Astrophysics, Carnegie Mellon University, 5000 Forbes Ave, Pittsburgh, PA 15213, USA }

\author[0000-0002-6021-8760]{Andrew B. Pace}
\affiliation{Department of Astronomy, University of Virginia, 530 McCormick Road, Charlottesville, VA 22904, USA}

\author{Jorge Pe\~narrubia}
\affiliation{Institute for Astronomy, University of Edinburgh, Royal Observatory, Blackford Hill, Edinburgh EH9 3HJ, UK}

\author[0009-0001-7781-6265]{Sofia L. Splawska}
\affiliation{McWilliams Center for Cosmology and Astrophysics, Carnegie Mellon University, 5000 Forbes Ave, Pittsburgh, PA 15213, USA }

\author[0000-0002-2732-9717]{Eduardo Vitral}\thanks{Royal Society Newton International Fellow}
\affiliation{Institute for Astronomy, University of Edinburgh, Royal Observatory, Blackford Hill, Edinburgh EH9 3HJ, UK}

\begin{abstract}
We fit a flexible double power-law (`$\alpha\beta\gamma$') model to the stellar density fields observed for the Milky Way's known dwarf spheroidal satellite galaxies.  We show that where standard criteria for model selection are decisive, the $\alpha\beta\gamma$ model is favored over the special case of the Plummer model, and also over the \sersic\ model and its special case, the exponential.  The Milky Way's dSph population exhibits a diverse range of stellar density profile shapes, as quantified by the values we infer for outer and inner power-law indices $\beta$ and $\gamma$.  Several of the most massive dSphs (e.g., Eridanus II, Fornax, Leo I, Leo II) have steeply-declining outer profiles, with $\beta\gtrsim 8$; others (e.g., Sextans, \booI) fade slowly, with $\beta\lesssim 4$.  The inner profiles of dSphs with stellar mass $\gtrsim 10^5 M_{\odot}$ are consistent with `cores' of uniform stellar density ($\gamma\approx 0$).  At lower masses the slopes of inner density profiles are poorly constrained, except in a few ultrafaint dSphs (e.g., Hercules, Ursa Major II) where we infer steep stellar cusps, with $\gamma\gtrsim 1.5$.  Owing to the $\alpha\beta\gamma$ model's flexibility, the inferred halflight radii and total stellar masses are significantly more uncertain than previous estimates, with halflight radii larger by up to an order of magnitude in some cases.  Finally, we demonstrate that allowing for flexibility in the shape of the stellar density profile is crucial for dSph mass modeling, where systematic errors associated with choice of stellar density profile can outweigh random errors in the observed velocity dispersions. 

\end{abstract}

\section{Introduction}

As the smallest, dimmest and most chemically primitive galaxies, the Milky Way's dwarf spheroidal\footnote{We use the term 'dwarf spheroidal galaxy (dSph)' to refer to a pressure-supported dwarf galaxy that has negligible cold gas and dormant star formation.  Included under this definition are the so-called `classical' dwarf spheroidals as well as 'ultrafaint' dwarf galaxies, which some authors distinguish according to luminosity.} satellites are well-established probes
of galaxy formation and dark matter.  Their stellar populations are resolved, with individual stars cataloged by flux, color, position, redshift, proper motion, chemical abundance, etc.  Conditioned on these quantities are inferences about astrophysical processes ranging from star formation to dark matter annihilation. 

One of the most basic properties inferred from dSph stellar catalogs is the spatial distribution of stars.  Most discoveries of new dSphs, as well as dynamical mass modeling of their stellar kinematics, are content to characterize stellar structural parameters (halflight radius, luminosity) based on the adoption of one of a few standard density profiles having fixed shape---e.g., exponential or \citet{plummer11} models \citep[e.g.,][and references therein]{simon19,read21}.  Yet systematic comparisons of different stellar density models usually demonstrate that some fit better than others, and in some cases the simplest models give poor fits to star count data.  

For example, \citet{ih95} find that \citet{king62} models fit the eight `classical'
dSphs significantly better than exponential models, but that even the King models tend to fail in outer regions, where they usually decline faster than the observed density profiles.  Beyond a `break' radius, the outer density profile often resembles a power law and/or \citet{sersic68} function, motivating Plummer and generalized exponential models as simple alternatives to the King profile.  Analyzing deep CFHT/Magellan MegaCam photometry that includes most of the known `ultrafaint' dSphs, \citet{munoz18} find that \sersic\ models tend to provide statistically better fits than exponential, Plummer and King models---a fact they attribute to the S\'{e}rsic model's greater flexibility.  Focusing on the Sextans dSph, \citet{cicuendez18} analyze DECam photometry and obtain statistically acceptable fits for all of the above models, finding a best fit using the King model.  On the other hand, examining CFHT/MegaCam and Dark Energy Survey observations of the Draco and Fornax dSphs, respectively, \citet{segall07} and \citet{wang19} fit all of the aforementioned models and find that \textit{none} give statistically acceptable fits, with even the best-fitting \sersic\ models failing to capture the observed inner and outer density profiles simultaneously.  \citet{valenciano25} obtain acceptable fits of Plummer, \sersic\ and Einasto models to deep \textit{HST} photometry of six ultrafaint dSphs \citep{richstein24}, with all but the Plummer model also capable of fitting relatively shallow \textit{Gaia} data for Fornax.

Some authors have used multi-component versions of standard surface brightness models to achieve greater flexibility.  In their analysis of dSph stellar kinematics, \citet{read19} model the stellar density profile as a sum of Plummer spheres.  Adopting the multi-Plummer model to characterize the stellar density profiles of all known Milky Way dSphs, \citet{moskowitz20} find that Fornax, Leo I, Leo II and Reticulum II all favor outer profiles that decline more steeply than the standard single Plummer profile.  \citet{jensen24} fit two-component exponential profiles to multi-dimensional \textit{Gaia} data, finding evidence for extended, faint outer populations in nine dSphs.  

All of the standard models named above carry strong assumptions about the behavior of dSph stellar density profiles.  King and Plummer models have central `cores' of uniform density.  At large radius, Plummer and exponential models decay at fixed rates.  King, \sersic\ and Einasto models all admit extra concentration parameters, but none has freedom to vary inner and outer slopes independently.  Even the multi-component versions asymptote to the inner and outer logarithmic slopes of the adopted basis function, giving flexibility only at intermediate radii.

Yet there is value in \textit{measuring}, rather than assuming, how dSph stellar density profiles behave at both small and large radius.  Indeed, departures from standard fitting functions observed in the outskirts of some dSphs have long fueled debate about their degree of tidal stripping due to the Milky Way \citep[e.g.,][]{ih95,martinez-delgado01,walcher03,majewski03,sohn07}, and hence about the validity of assuming an equilibrium state when applying dynamical models \citep[e.g.,][]{oh95,piatek95,read06}.  On the theoretical side, \citet{penarrubia09} use controlled N-body experiments to demonstrate that, once equilibrium is re-established after tidal mass loss events, the outer stellar density profiles of dSph analogs embedded within standard cold dark matter halos tend to evolve toward shallower, Plummer-like power laws.  Subsequent simulations by \citet{errani24} examine how the tidal evolution of a dSph satellite's stellar component depends sensitively on the \text{inner} stellar density profile, which dictates the degree to which stars populate the most bound energy states.  Specifically, \citet{errani24} show that if stars follow the `cuspy' central density profile of their host dark matter halo, then even the most severe tidal stripping will leave a bound luminous remnant, or `microgalaxy'.  Moreover, in a companion paper, \citet{splawska26} explore how dynamical mass estimators, and the scaling relations derived therefrom, are sensitive to the shapes of the stellar density profiles.

The shapes of dSph stellar density profiles may even be diagnostic of the dark matter distribution independently of stellar-kinematic observations.  For example, given the `cuspy' dark matter density profile expected for standard cold dark matter halos \citep{navarro97}, there is no physical stellar phase-space distribution function corresponding to a spherically-symmetric and isotropic stellar configuration with a `core' of uniform central density \citep{sanchez-almeida23}.  Based on this fact, \citet{sanchez-almeida24} argue that the cored model they fit to the (stacked) stellar density profiles observed for several ultrafaint dSphs disfavors host halos made of standard cold dark matter.  While this conclusion may rest precariously on the degree to which a perfectly cored density profile can be inferred from finite data \citep{hakkinen25,valenciano25}, the shapes of dSph stellar density profiles clearly have implications for dynamical models.

Here we use a relatively flexible, double power-law model to measure the behaviors of dSph stellar density profiles at both small and large radius.  We perform a homogeneous analysis of public catalog-level data from six sky surveys and make all results publicly available.    We compare results to those obtained using standard Plummer and \sersic\ models, applying quantitative criteria for model selection.  We examine the distribution of luminous structural parameters that characterize the dSph population under our flexible model.  Finally, we consider implications for mass modeling, as the stellar density profile enters directly into, e.g., the Jeans equations \citep{bt08}.  

\section{Data}

We consider \ngal\ dwarf spheroidal galaxies that are companions of the Milky Way, as listed in the `dwarf\_mw' table of the Local Volume Database, v1.0.6 \citep[LVDB][]{pace24}; from this table we exclude only the Magellanic Clouds and Sagittarius.  The first column of Table \ref{tab:sbfit_table} lists names of the \ngood\ galaxies for which we report results here.  Results for the remaining \nbad\ galaxies---\badnames---do not pass the quality criteria discussed in Section \ref{sec:analysis}, but for completeness are included in the available electronic data products (see Appendix).   

\begin{deluxetable*}{llllllllllllllll} 
\tablewidth{0pt} 
\tablecaption{Summary of Fitting Results, $\alpha\beta\gamma$ model\tablenotemark{$\dagger$} \label{tab:sbfit_table}} 
\tablehead{\colhead{$^{\rm Galaxy}_{\rm \qquad\; Catalog}$}&\colhead{$^{\alpha_0\,\mathrm{[deg]}}_{\delta_0\,\mathrm{[deg]}}$}&\colhead{$\alpha$}&\colhead{$\beta$}&\colhead{$\gamma$}&\colhead{$\log_{10}\bigl [\frac{r_s}{\mathrm{arcmin}}\bigr ]$}&\colhead{$^{\epsilon\equiv1-b/a}_{\theta\,\mathrm{[deg]}}$}&\colhead{$\log_{10}N_{\rm mem}^{\rm field}$}&\colhead{$\log_{10}\bigl [\frac{R_{\rm h}}{\mathrm{pc}}\bigr ]$}&\colhead{$\log_{10}\bigl [\frac{M_{\rm mem}^{\rm tot}}{M_{\odot}}\bigr ]$}} 
\startdata 
$^{\rm Aquarius\, II}_{ \rm \qquad\; DECaLS}$ & $^{338.481 \pm 0.015}_{-009.319 \pm 0.005}$ & $1.66 \pm 0.65$ & $5.74 \pm 1.87$ & $0.22 \pm 0.72$ & $1.04 \pm 0.29$ & $^{0.65 \pm 0.17}_{-78 \pm 29}$ & $2.19 \pm 0.10$ & $2.22 \pm 0.27\dagger$ & $3.98 \pm 0.15$ & $$\\ 
$^{\rm Bootes\, I}_{ \rm \qquad\; DECaLS}$ & $^{210.021 \pm 0.004}_{+014.515 \pm 0.006}$ & $2.27 \pm 0.53$ & $3.40 \pm 0.55$ & $-0.43 \pm 0.41$ & $0.89 \pm 0.10$ & $^{0.41 \pm 0.03}_{+10 \pm 3}$ & $3.42 \pm 0.03$ & $2.84 \pm 0.50\dagger$ & $5.05 \pm 0.16$ & $$\\ 
$^{\rm Bootes\, II}_{ \rm \qquad\; DECaLS}$ & $^{209.525 \pm 0.006}_{+012.856 \pm 0.004}$ & $2.00 \pm 0.61$ & $6.93 \pm 1.78$ & $0.31 \pm 0.69$ & $0.74 \pm 0.20$ & $^{0.27 \pm 0.13}_{-67 \pm 63}$ & $2.13 \pm 0.08$ & $1.55 \pm 0.14\dagger$ & $3.03 \pm 0.10$ & $$\\ 
$^{\rm Bootes\, IV}_{ \rm \qquad\; DECaLS}$ & $^{233.691 \pm 0.016}_{+043.721 \pm 0.016}$ & $1.54 \pm 0.69$ & $5.02 \pm 1.81$ & $1.37 \pm 0.70$ & $1.48 \pm 0.31$ & $^{0.39 \pm 0.20}_{-24 \pm 32}$ & $2.32 \pm 0.19$ & $2.97 \pm 0.51\dagger$ & $4.64 \pm 0.30\dagger$ & $$\\ 
$^{\rm Bootes\, V}_{ \rm \qquad\; DECaLS}$ & $^{213.912 \pm 0.002}_{+032.913 \pm 0.002}$ & $1.52 \pm 0.66$ & $5.73 \pm 1.92$ & $0.67 \pm 0.82$ & $0.18 \pm 0.35$ & $^{0.59 \pm 0.19}_{+38 \pm 17}$ & $1.41 \pm 0.12$ & $1.32 \pm 0.31\dagger$ & $3.11 \pm 0.18$ & $$\\ 
$^{\rm Canes\, Venatici\, I}_{ \rm \qquad\; DECaLS}$ & $^{202.005 \pm 0.005}_{+033.551 \pm 0.003}$ & $1.55 \pm 0.44$ & $8.01 \pm 1.37$ & $-0.08 \pm 0.52$ & $1.13 \pm 0.12$ & $^{0.46 \pm 0.04}_{+77 \pm 3}$ & $2.95 \pm 0.02$ & $2.51 \pm 0.02$ & $5.23 \pm 0.03$ & $$\\ 
$^{\rm Canes\, Venatici\, II}_{ \rm \qquad\; DECaLS}$ & $^{194.293 \pm 0.003}_{+034.319 \pm 0.004}$ & $1.91 \pm 0.62$ & $7.13 \pm 1.77$ & $0.27 \pm 0.72$ & $0.44 \pm 0.21$ & $^{0.39 \pm 0.16}_{-00 \pm 18}$ & $1.78 \pm 0.08$ & $1.76 \pm 0.12\dagger$ & $3.93 \pm 0.09$ & $$\\ 
$^{\rm Carina}_{ \rm \qquad\; Gaia}$ & $^{100.388 \pm 0.006}_{-050.960 \pm 0.003}$ & $2.50 \pm 0.47$ & $5.71 \pm 0.94$ & $0.43 \pm 0.40$ & $1.12 \pm 0.08$ & $^{0.36 \pm 0.02}_{+60 \pm 2}$ & $3.08 \pm 0.01$ & $2.38 \pm 0.03$ & $5.63 \pm 0.02$ & $$\\ 
$^{\rm Carina\, II}_{ \rm \qquad\; DELVE}$ & $^{114.155 \pm 0.023}_{-057.977 \pm 0.014}$ & $1.14 \pm 0.58$ & $4.14 \pm 1.61$ & $0.05 \pm 0.60$ & $1.18 \pm 0.30\dagger$ & $^{0.17 \pm 0.10}_{+42 \pm 31}$ & $3.20 \pm 0.09$ & $2.52 \pm 0.48\dagger$ & $4.33 \pm 0.22$ & $$\\ 
$^{\rm Carina\, III}_{ \rm \qquad\; DELVE}$ & $^{114.546 \pm 0.016}_{-057.940 \pm 0.011}$ & $2.48 \pm 0.49$ & $5.73 \pm 1.79$ & $-0.57 \pm 0.38$ & $1.48 \pm 0.08$ & $^{0.54 \pm 0.07}_{+81 \pm 19}$ & $3.03 \pm 0.09$ & $2.16 \pm 0.36\dagger$ & $3.88 \pm 0.26\dagger$ & $$\\ 
$^{\rm Centaurus\, I}_{ \rm \qquad\; DELVE}$ & $^{189.591 \pm 0.004}_{-040.898 \pm 0.004}$ & $2.00 \pm 0.67$ & $6.44 \pm 1.93$ & $0.30 \pm 0.83$ & $0.56 \pm 0.27$ & $^{0.32 \pm 0.13}_{+37 \pm 18}$ & $2.12 \pm 0.11$ & $1.84 \pm 0.22\dagger$ & $4.36 \pm 0.14$ & $$\\ 
$^{\rm Cetus\, II}_{ \rm \qquad\; DECaLS}$ & $^{019.455 \pm 0.010}_{-017.442 \pm 0.010}$ & $1.60 \pm 0.72$ & $4.91 \pm 1.90$ & $0.57 \pm 0.86$ & $0.83 \pm 0.31$ & $^{0.63 \pm 0.12}_{+44 \pm 9}$ & $1.90 \pm 0.13$ & $1.51 \pm 0.46\dagger$ & $2.63 \pm 0.23\dagger$ & $$\\ 
$^{\rm Columba\, I}_{ \rm \qquad\; DECaLS}$ & $^{082.853 \pm 0.007}_{-028.037 \pm 0.006}$ & $1.85 \pm 0.64$ & $6.51 \pm 1.84$ & $0.23 \pm 0.72$ & $0.61 \pm 0.24$ & $^{0.23 \pm 0.16}_{+40 \pm 41}$ & $1.75 \pm 0.10$ & $2.11 \pm 0.20\dagger$ & $3.96 \pm 0.13$ & $$\\ 
$^{\rm Coma\, Berenices}_{ \rm \qquad\; DECaLS}$ & $^{186.746 \pm 0.005}_{+023.913 \pm 0.003}$ & $1.95 \pm 0.62$ & $7.20 \pm 1.70$ & $0.94 \pm 0.54$ & $1.17 \pm 0.17$ & $^{0.36 \pm 0.06}_{-64 \pm 6}$ & $2.75 \pm 0.04$ & $1.86 \pm 0.07\dagger$ & $3.68 \pm 0.06$ & $$\\ 
$^{\rm Crater\, II}_{ \rm \qquad\; Gaia}$ & $^{177.310 \pm 0.023}_{-018.397 \pm 0.021}$ & $2.57 \pm 0.34$ & $8.37 \pm 1.24$ & $-0.59 \pm 0.35$ & $1.57 \pm 0.07$ & $^{0.08 \pm 0.06}_{+38 \pm 53}$ & $2.30 \pm 0.04$ & $2.89 \pm 0.03$ & $4.85 \pm 0.04$ & $$\\ 
$^{\rm Draco}_{ \rm \qquad\; DECaLS}$ & $^{260.065 \pm 0.003}_{+057.921 \pm 0.001}$ & $2.56 \pm 0.36$ & $5.47 \pm 0.49$ & $0.20 \pm 0.24$ & $1.09 \pm 0.04$ & $^{0.32 \pm 0.01}_{-89 \pm 81}$ & $3.97 \pm 0.01$ & $2.30 \pm 0.01$ & $5.45 \pm 0.02$ & $$\\ 
$^{\rm Draco\, II}_{ \rm \qquad\; DECaLS}$ & $^{238.219 \pm 0.011}_{+064.565 \pm 0.005}$ & $1.47 \pm 0.66$ & $5.45 \pm 1.92$ & $0.65 \pm 0.75$ & $0.76 \pm 0.33$ & $^{0.22 \pm 0.15}_{+41 \pm 47}$ & $2.10 \pm 0.11$ & $1.39 \pm 0.36\dagger$ & $2.56 \pm 0.18$ & $$\\ 
$^{\rm Eridanus\, II}_{ \rm \qquad\; DECaLS}$ & $^{056.087 \pm 0.003}_{-043.534 \pm 0.001}$ & $2.44 \pm 0.42$ & $8.40 \pm 1.17$ & $0.05 \pm 0.49$ & $0.60 \pm 0.08$ & $^{0.42 \pm 0.04}_{+79 \pm 4}$ & $2.51 \pm 0.03$ & $2.27 \pm 0.02$ & $5.03 \pm 0.03$ & $$\\ 
$^{\rm Eridanus\, IV}_{ \rm \qquad\; DELVE}$ & $^{076.433 \pm 0.010}_{-009.512 \pm 0.007}$ & $1.82 \pm 0.63$ & $6.50 \pm 1.86$ & $0.19 \pm 0.68$ & $0.90 \pm 0.23$ & $^{0.49 \pm 0.14}_{+64 \pm 16}$ & $2.01 \pm 0.09$ & $1.89 \pm 0.19\dagger$ & $3.60 \pm 0.13$ & $$\\ 
$^{\rm Fornax}_{ \rm \qquad\; DECaLS}$ & $^{039.958 \pm 0.001}_{-034.515 \pm 0.001}$ & $1.80 \pm 0.10$ & $9.68 \pm 0.33$ & $0.18 \pm 0.13$ & $1.57 \pm 0.02$ & $^{0.31 \pm 0.00}_{+43 \pm 0}$ & $4.57 \pm 0.00$ & $2.76 \pm 0.01$ & $7.13 \pm 0.01$ & $$\\ 
$^{\rm Grus\, I}_{ \rm \qquad\; DECaLS}$ & $^{344.178 \pm 0.009}_{-050.179 \pm 0.009}$ & $1.62 \pm 0.68$ & $5.49 \pm 1.93$ & $1.07 \pm 0.86$ & $0.90 \pm 0.34\dagger$ & $^{0.30 \pm 0.17}_{-29 \pm 30}$ & $2.00 \pm 0.14$ & $2.20 \pm 0.39\dagger$ & $4.02 \pm 0.21$ & $$\\ 
$^{\rm Grus\, II}_{ \rm \qquad\; DECaLS}$ & $^{331.023 \pm 0.010}_{-046.434 \pm 0.007}$ & $1.90 \pm 0.61$ & $6.51 \pm 1.87$ & $0.25 \pm 0.63$ & $0.99 \pm 0.20$ & $^{0.10 \pm 0.08}_{-06 \pm 53}$ & $2.53 \pm 0.07$ & $2.00 \pm 0.20\dagger$ & $3.69 \pm 0.12$ & $$\\ 
$^{\rm Hercules}_{ \rm \qquad\; DECaLS}$ & $^{247.759 \pm 0.008}_{+012.786 \pm 0.002}$ & $1.10 \pm 0.67$ & $3.87 \pm 1.24$ & $1.46 \pm 0.73$ & $1.32 \pm 0.40$ & $^{0.64 \pm 0.08}_{-76 \pm 4}$ & $2.73 \pm 0.09$ & $2.86 \pm 0.62\dagger$ & $4.93 \pm 0.26\dagger$ & $$\\ 
$^{\rm Horologium\, I}_{ \rm \qquad\; DECaLS}$ & $^{043.872 \pm 0.005}_{-054.118 \pm 0.003}$ & $1.99 \pm 0.56$ & $6.51 \pm 1.75$ & $-0.10 \pm 0.58$ & $0.45 \pm 0.18$ & $^{0.18 \pm 0.11}_{+41 \pm 31}$ & $2.11 \pm 0.05$ & $1.62 \pm 0.10\dagger$ & $3.64 \pm 0.08$ & $$\\ 
$^{\rm Horologium\, II}_{ \rm \qquad\; DECaLS}$ & $^{049.128 \pm 0.007}_{-050.008 \pm 0.004}$ & $2.10 \pm 0.62$ & $7.20 \pm 1.77$ & $0.21 \pm 0.74$ & $0.45 \pm 0.20$ & $^{0.21 \pm 0.14}_{-41 \pm 60}$ & $1.62 \pm 0.10$ & $1.54 \pm 0.18\dagger$ & $3.13 \pm 0.16$ & $$\\ 
$^{\rm Hydrus\, I}_{ \rm \qquad\; DELVE}$ & $^{037.413 \pm 0.024}_{-079.305 \pm 0.003}$ & $2.11 \pm 0.52$ & $5.76 \pm 1.48$ & $-0.16 \pm 0.46$ & $0.97 \pm 0.13$ & $^{0.23 \pm 0.05}_{-78 \pm 84}$ & $3.14 \pm 0.03$ & $1.74 \pm 0.04$ & $3.78 \pm 0.04$ & $$\\ 
$^{\rm Leo\, I}_{ \rm \qquad\; PS1}$ & $^{152.116 \pm 0.002}_{+012.308 \pm 0.001}$ & $2.49 \pm 0.35$ & $8.01 \pm 0.97$ & $-0.04 \pm 0.54$ & $0.87 \pm 0.06$ & $^{0.35 \pm 0.02}_{+85 \pm 4}$ & $3.26 \pm 0.01$ & $2.42 \pm 0.02$ & $5.90 \pm 0.02$ & $$\\ 
$^{\rm Leo\, II}_{ \rm \qquad\; DECaLS}$ & $^{168.364 \pm 0.001}_{+022.152 \pm 0.001}$ & $2.68 \pm 0.27$ & $8.78 \pm 0.78$ & $-0.12 \pm 0.36$ & $0.66 \pm 0.04$ & $^{0.03 \pm 0.02}_{+37 \pm 44}$ & $3.20 \pm 0.01$ & $2.20 \pm 0.01$ & $5.72 \pm 0.01$ & $$\\ 
$^{\rm Leo\, IV}_{ \rm \qquad\; DECaLS}$ & $^{173.244 \pm 0.005}_{-000.545 \pm 0.005}$ & $1.75 \pm 0.66$ & $5.30 \pm 1.93$ & $0.02 \pm 0.69$ & $0.52 \pm 0.28$ & $^{0.19 \pm 0.12}_{-28 \pm 37}$ & $2.07 \pm 0.10$ & $2.11 \pm 0.31\dagger$ & $4.18 \pm 0.16$ & $$\\ 
$^{\rm Leo\, V}_{ \rm \qquad\; DELVE}$ & $^{172.783 \pm 0.004}_{+002.218 \pm 0.001}$ & $1.37 \pm 0.67$ & $4.72 \pm 1.73$ & $1.01 \pm 0.81$ & $0.38 \pm 0.36$ & $^{0.66 \pm 0.21}_{-80 \pm 74}$ & $1.51 \pm 0.13$ & $1.81 \pm 0.46\dagger$ & $3.79 \pm 0.22$ & $$\\ 
$^{\rm Leo\, VI}_{ \rm \qquad\; DELVE}$ & $^{171.065 \pm 0.024}_{+024.829 \pm 0.007}$ & $1.86 \pm 0.68$ & $4.10 \pm 0.93$ & $0.02 \pm 0.60$ & $1.51 \pm 0.12$ & $^{0.60 \pm 0.08}_{-45 \pm 6}$ & $2.73 \pm 0.10$ & $3.04 \pm 0.50\dagger$ & $4.89 \pm 0.30\dagger$ & $$\\ 
$^{\rm Leo\, Minor\, I}_{ \rm \qquad\; DECaLS}$ & $^{164.263 \pm 0.004}_{+028.876 \pm 0.003}$ & $2.04 \pm 0.64$ & $6.83 \pm 1.87$ & $0.06 \pm 0.72$ & $0.27 \pm 0.23$ & $^{0.18 \pm 0.14}_{+08 \pm 50}$ & $1.56 \pm 0.10$ & $1.43 \pm 0.21\dagger$ & $3.01 \pm 0.15$ & $$\\ 
$^{\rm Pegasus\, IV}_{ \rm \qquad\; DECaLS}$ & $^{328.540 \pm 0.003}_{+026.621 \pm 0.003}$ & $1.89 \pm 0.63$ & $7.00 \pm 1.77$ & $0.43 \pm 0.72$ & $0.43 \pm 0.22$ & $^{0.21 \pm 0.14}_{-30 \pm 36}$ & $1.91 \pm 0.08$ & $1.55 \pm 0.15\dagger$ & $3.63 \pm 0.10$ & $$\\ 
$^{\rm Phoenix\, II}_{ \rm \qquad\; DECaLS}$ & $^{354.995 \pm 0.004}_{-054.406 \pm 0.003}$ & $2.04 \pm 0.64$ & $7.16 \pm 1.78$ & $1.08 \pm 0.73$ & $0.56 \pm 0.21$ & $^{0.18 \pm 0.12}_{-11 \pm 34}$ & $1.80 \pm 0.08$ & $1.58 \pm 0.17\dagger$ & $3.38 \pm 0.14$ & $$\\ 
$^{\rm Pictor\, I}_{ \rm \qquad\; DECaLS}$ & $^{070.943 \pm 0.004}_{-050.286 \pm 0.003}$ & $1.46 \pm 0.67$ & $4.91 \pm 1.90$ & $0.50 \pm 0.79$ & $0.31 \pm 0.37\dagger$ & $^{0.47 \pm 0.17}_{+41 \pm 20}$ & $1.65 \pm 0.11$ & $1.71 \pm 0.42\dagger$ & $3.55 \pm 0.20$ & $$\\ 
$^{\rm Pictor\, II}_{ \rm \qquad\; DELVE}$ & $^{101.181 \pm 0.005}_{-059.884 \pm 0.004}$ & $1.44 \pm 0.70$ & $4.81 \pm 1.88$ & $0.76 \pm 0.82$ & $0.24 \pm 0.40\dagger$ & $^{0.12 \pm 0.09}_{+07 \pm 38}$ & $2.18 \pm 0.10$ & $1.31 \pm 0.30\dagger$ & $3.38 \pm 0.16$ & $$\\ 
$^{\rm Pisces\, II}_{ \rm \qquad\; DECaLS}$ & $^{344.631 \pm 0.003}_{+005.953 \pm 0.002}$ & $1.83 \pm 0.67$ & $6.14 \pm 1.95$ & $0.23 \pm 0.80$ & $0.29 \pm 0.27$ & $^{0.45 \pm 0.16}_{+68 \pm 57}$ & $1.63 \pm 0.11$ & $1.76 \pm 0.31\dagger$ & $3.86 \pm 0.16$ & $$\\ 
$^{\rm Reticulum\, II}_{ \rm \qquad\; DECaLS}$ & $^{053.924 \pm 0.006}_{-054.048 \pm 0.002}$ & $1.94 \pm 0.48$ & $6.95 \pm 1.40$ & $-0.35 \pm 0.44$ & $0.94 \pm 0.11$ & $^{0.56 \pm 0.02}_{+66 \pm 2}$ & $2.86 \pm 0.02$ & $1.56 \pm 0.03$ & $3.60 \pm 0.04$ & $$\\ 
$^{\rm Sculptor}_{ \rm \qquad\; DELVE}$ & $^{015.020 \pm 0.001}_{-033.714 \pm 0.001}$ & $2.73 \pm 0.21$ & $4.50 \pm 0.08$ & $-0.15 \pm 0.18$ & $1.06 \pm 0.02$ & $^{0.27 \pm 0.01}_{-87 \pm 1}$ & $4.58 \pm 0.00$ & $2.45 \pm 0.01$ & $6.08 \pm 0.02$ & $$\\ 
$^{\rm Segue\, 1}_{ \rm \qquad\; DECaLS}$ & $^{151.753 \pm 0.006}_{+016.075 \pm 0.004}$ & $1.54 \pm 0.67$ & $5.80 \pm 1.96$ & $0.64 \pm 0.71$ & $0.87 \pm 0.29$ & $^{0.32 \pm 0.13}_{+72 \pm 49}$ & $2.29 \pm 0.09$ & $1.47 \pm 0.30\dagger$ & $2.79 \pm 0.16$ & $$\\ 
$^{\rm Segue\, 2}_{ \rm \qquad\; DECaLS}$ & $^{034.821 \pm 0.003}_{+020.164 \pm 0.005}$ & $2.12 \pm 0.57$ & $6.64 \pm 1.70$ & $0.16 \pm 0.60$ & $0.79 \pm 0.16$ & $^{0.34 \pm 0.09}_{-10 \pm 9}$ & $2.39 \pm 0.05$ & $1.55 \pm 0.07$ & $3.23 \pm 0.08$ & $$\\ 
$^{\rm Sextans}_{ \rm \qquad\; DELVE}$ & $^{153.253 \pm 0.004}_{-001.631 \pm 0.003}$ & $2.00 \pm 0.39$ & $4.26 \pm 0.33$ & $-0.37 \pm 0.38$ & $1.25 \pm 0.05$ & $^{0.25 \pm 0.01}_{+62 \pm 2}$ & $4.13 \pm 0.01$ & $2.76 \pm 0.03$ & $5.67 \pm 0.06$ & $$\\ 
$^{\rm Triangulum\, II}_{ \rm \qquad\; PS1}$ & $^{033.327 \pm 0.011}_{+036.187 \pm 0.008}$ & $1.62 \pm 0.66$ & $5.51 \pm 1.90$ & $0.44 \pm 0.74$ & $0.79 \pm 0.29$ & $^{0.28 \pm 0.18}_{+51 \pm 49}$ & $1.91 \pm 0.15$ & $1.53 \pm 0.38\dagger$ & $3.09 \pm 0.23\dagger$ & $$\\ 
$^{\rm Tucana\, II}_{ \rm \qquad\; DECaLS}$ & $^{342.977 \pm 0.025}_{-058.587 \pm 0.009}$ & $1.54 \pm 0.67$ & $6.01 \pm 1.92$ & $0.88 \pm 0.69$ & $1.48 \pm 0.29$ & $^{0.38 \pm 0.10}_{-71 \pm 30}$ & $2.75 \pm 0.08$ & $2.38 \pm 0.25\dagger$ & $3.94 \pm 0.16$ & $$\\ 
$^{\rm Tucana\, III}_{ \rm \qquad\; DECaLS}$ & $^{359.142 \pm 0.013}_{-059.601 \pm 0.007}$ & $2.00 \pm 0.63$ & $6.73 \pm 1.90$ & $0.09 \pm 0.70$ & $0.84 \pm 0.20$ & $^{0.18 \pm 0.12}_{-41 \pm 46}$ & $2.23 \pm 0.10$ & $1.45 \pm 0.20\dagger$ & $2.77 \pm 0.12$ & $$\\ 
$^{\rm Tucana\, IV}_{ \rm \qquad\; DES}$ & $^{000.697 \pm 0.026}_{-060.827 \pm 0.018}$ & $1.24 \pm 0.65$ & $4.64 \pm 1.80$ & $0.41 \pm 0.74$ & $1.27 \pm 0.34\dagger$ & $^{0.44 \pm 0.15}_{-08 \pm 14}$ & $2.63 \pm 0.12$ & $2.38 \pm 0.48\dagger$ & $3.71 \pm 0.24\dagger$ & $$\\ 
$^{\rm Tucana\, V}_{ \rm \qquad\; DECaLS}$ & $^{354.343 \pm 0.005}_{-063.266 \pm 0.003}$ & $1.54 \pm 0.70$ & $5.91 \pm 1.95$ & $1.12 \pm 0.80$ & $0.59 \pm 0.34$ & $^{0.38 \pm 0.16}_{+39 \pm 23}$ & $1.82 \pm 0.10$ & $1.46 \pm 0.32\dagger$ & $2.98 \pm 0.18$ & $$\\ 
$^{\rm Ursa\, Major\, I}_{ \rm \qquad\; DECaLS}$ & $^{158.663 \pm 0.025}_{+051.917 \pm 0.011}$ & $1.25 \pm 0.71$ & $3.58 \pm 1.05$ & $0.08 \pm 0.70$ & $1.12 \pm 0.28\dagger$ & $^{0.72 \pm 0.05}_{+62 \pm 3}$ & $2.72 \pm 0.12$ & $3.07 \pm 0.70\dagger$ & $4.73 \pm 0.32\dagger$ & $$\\ 
$^{\rm Ursa\, Major\, II}_{ \rm \qquad\; DECaLS}$ & $^{132.874 \pm 0.004}_{+063.133 \pm 0.001}$ & $2.53 \pm 0.46$ & $3.64 \pm 1.56$ & $1.77 \pm 0.13$ & $1.55 \pm 0.16$ & $^{0.59 \pm 0.04}_{-78 \pm 2}$ & $3.22 \pm 0.07$ & $2.49 \pm 0.59\dagger$ & $4.10 \pm 0.22$ & $$\\ 
$^{\rm Ursa\, Minor}_{ \rm \qquad\; DECaLS}$ & $^{227.260 \pm 0.005}_{+067.226 \pm 0.002}$ & $2.89 \pm 0.14$ & $4.57 \pm 0.11$ & $0.14 \pm 0.13$ & $1.27 \pm 0.02$ & $^{0.54 \pm 0.01}_{+50 \pm 0}$ & $4.21 \pm 0.01$ & $2.48 \pm 0.02$ & $5.72 \pm 0.06$ & $$\\ 
$^{\rm Willman\, 1}_{ \rm \qquad\; DECaLS}$ & $^{162.345 \pm 0.004}_{+051.051 \pm 0.002}$ & $1.25 \pm 0.62$ & $5.53 \pm 1.87$ & $0.43 \pm 0.76$ & $0.63 \pm 0.34$ & $^{0.44 \pm 0.07}_{+77 \pm 16}$ & $2.37 \pm 0.05$ & $1.44 \pm 0.23\dagger$ & $3.21 \pm 0.17$ & $$\\ 
\enddata 
\tablenotetext{\dagger}{indicates logarithmic quantities where the corresponding linear quantity is unresolved, with standard deviation larger than the mean} 
\end{deluxetable*}

For each galaxy we adopt public, catalog-level photometric data from up to six large-scale sky surveys: Sloan Digital Sky Survey Data Release 9 (SDSS DR9, \citealt{sdssdr9}), Pan-STARRS Data Release 1 (PS1, \citealt{ps1}), Dark Energy Survey Data Release 2 (DES DR2, \citealt{desdr2}), Dark Energy Camera Legacy Survey Data Release 9 (DECaLS DR9, \citealt{dey19}), DECam Local Volume Exploration Survey Data Release 3 (DELVE DR3\footnote{available at https://datalab.noirlab.edu/data/delve}, \citealt{drlica-wagner21,drlica-wagner22}), and \textit{Gaia} Data Release 3 (DR3, \citealt{gaia16,gaiadr3}).  Not all of the adopted survey catalogs are independent.  For example, DECaLS combines observations acquired using the Dark Energy Camera (DECam) at the 4m Blanco telescope with data (also acquired with DECam) from the Dark Energy Survey.  DELVE combines DES and DECaLS imaging data with additional DECam observations, adding a wide-area survey at high Galactic latitude and deeper observations targeting the periphery of the Magellanic Clouds.  While DELVE uses the same pipelines as DES for image processing and photometry \citep{morganson18}, DECaLS uses an independent pipeline \citep{dey19}.  As a result, the DECaLS, DES and DELVE catalogs are all correlated due to (at least partial) use of the same source images, and DES/DELVE are further correlated by use of the same processing pipelines.

For each survey and each target galaxy, we download catalog-level information for all sources projected within a square field that is centered on the galaxy and has side length $20 R_{\rm half}$, where $R_{\rm half}$ is the projected halflight radius.  For this purpose we adopt nominal values for the galaxy center and $R_{\rm half}$ from the LVDB.

We impose survey-specific filters to select likely point sources and discard likely extended sources.  For SDSS we require \texttt{mode}=1 and \texttt{type}=6.  For PS1 we require $\textrm{\texttt{rpsfmag}}-\textrm{\texttt{rkronmag}}<0.05$.  For DES and DELVE we require $0\leq \textrm{\texttt{EXTENDED\_CLASS\_G}}\leq1$, where the \texttt{EXTENDED\_CLASS} parameter is defined as in \citet{desdr2}.  For DECaLS we require \texttt{type}=PSF.  For \textit{Gaia} we require $\textrm{\texttt{astrometric\_excess\_noise\_sig}}<2$, renormalized unit weight error $\textrm{\texttt{RUWE}}<1.4$, $\textrm{\texttt{parallax}}<3\times \textrm{\texttt{parallax\_error}}$, and we discard sources that are cross-listed in Gaia's variable source catalog of active galactic nuclei.

\subsection{Color-magnitude selection}
\label{subsec:cmd_filter}
We use the dust maps of \citet{sfd98}, as recalibrated by \citet{schlafly11} and implemented in the \texttt{dustmaps} software package \citep{green18}, to estimate and apply filter-specific extinction corrections to all the survey-catalogued photometry.  We use extinction-corrected magnitudes and colors to select candidate dwarf galaxy member stars following methodology that is similar to the one described by \citet{an24} and summarized below.  For all surveys except  Gaia, we use $g$- and $r$-band photometry.  For \textit{Gaia} we use $G$-band magnitude and the $BP-RP$ color derived from low-resolution spectroscopy.

First, for each galaxy/survey combination, we compute the distance of each star, in color/magnitude space, from the best-fitting (by eye) theoretical isochrone.  We use the MIniMist INTerpolation software package (\texttt{Minimint}, \citealt{minimint}) to interpolate the MESA Isochrones and Stellar Tracks (MIST, \citealt{dotter16}) at the galaxy's mean metallicity (as tabulated by the LVDB; where unavailable, as for some ultra-faint systems, we assume [Fe/H]$=-2$) and old age (typically $10-13$ Gyr).  With these constraints, in some cases we manually shift the isochrone color by up to $\sim 0.1$ magnitude to improve agreement with photometry of the likeliest member stars (those within $1 R_{\rm half}$ of the center).  We consider stars within $\epsilon_m$ magnitudes of the isochrone to be candidate members, where we take $\epsilon_m$ to be the quadrature sum of the propagated (cataloged) uncertainty in the star's color and a value of $0.25$ mag that accounts for intrinsic width of the member sequence.

Second, we use a data-driven procedure to identify candidate member stars.  For each galaxy/survey combination, we discretize the color/magnitude plane into a $50\times 50$ grid of equally-spaced cells.  Within each cell we count the numbers of stars projected inside the published projected halflight radius (as tabulated by the LVDB) vs outside the circle of radius $3R_{\rm half}$.  We mask pixels for which the ratio of inner to outer stars falls below a specified percentile, which by default is the $75^{\rm th}$ but we take liberty to adjust as high as $95^{\rm th}$ to exclude cells that are clearly far from the sequence of galaxy members.  We then use the \texttt{area\_opening} algorithm (with \texttt{threshold}=16 and \texttt{connectivity}=1) from the Python-based \texttt{scikit-image} package \citep{scikit-image} to smooth the binary image of masked/unmasked cells.  Stars within cells that remain unmasked after this procedure are considered candidate members.  

For subsequent analysis we use the samples of candidate members that are passed by \textit{either} of the above procedures.  These samples include all the stars that would pass a standard isochrone-based color/magnitude filter, plus stars within regions of color/magnitude space that happen to contain relatively large fractions of stars within $1R_{\rm half}$.  Finally we impose bright-end and faint-end magnitude limits based on completeness limits reported by each survey.  For SDSS and PS1 we require $r\leq 22$; for DES, DECaLS and DELVE $r\leq 23.4$, and for \textit{Gaia} $G\leq 20.5$.  At the bright end we require $r\geq 16$ and/or $G\geq 16$.  

\begin{figure*}
\includegraphics[width=\textwidth,trim=0.in 0.3in 0.5in 0in, clip]{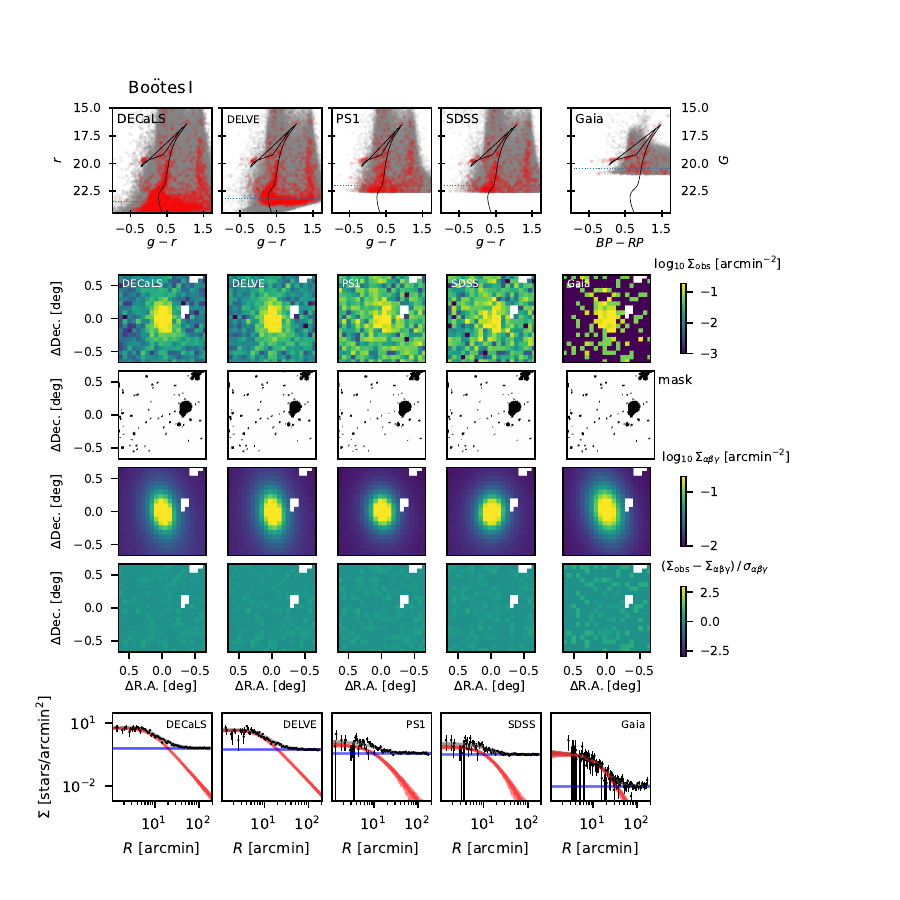}
    \caption{Illustration of analysis using \booI\ as an example.\textit{Top:} Color-magnitude diagrams of all point sources within $2^{\circ}$ of \booI\, from the denoted sky survey catalogs.  Red markers indicate sources within the (previously-published) 2D halflight radius.  Solid curves are isochrones corresponding to \boo\ I's (previously-published) age/metallicity; horizontal dotted lines denote the adopted magnitude limit.  \textit{Second row:} Surface density field of color/magnitude-selected sources, rebinned (as in rows 4 and 5) from $250\times 250$ pix$^2$ to $50\times 50$ pix$^2$ for display purposes.  \textit{Third row:} Pixel masks (original $250\times 250$ pix$^{2}$ grid) used when fitting the stellar density field, for the denoted surveys, with masked pixels appearing in black.  Masks account for survey footprint, presence of other known nearby galaxies and/or star clusters, and artifacts from bright sources.  \textit{Fourth row:} Best-fitting $\alpha\beta\gamma$ models of the stellar density field depicted in the second row.  \textit{Fifth row:} Difference between observed stellar surface density and best-fitting $\alpha\beta\gamma$ model, normalized by $1\sigma$ interval in the posterior probability distribution function.  \textit{Bottom:} Stellar surface density profile obtained using catalogs from in the indicated surveys, binned according to elliptical radius (based on previously-published ellipticity and position angle).  Overplotted in gray is the  $\alpha\beta\gamma$ model, decomposed into contributions from \booI\ members (red) and Galactic foreground (blue); envelopes enclose the central 68\% of the posterior probability distribution.   \label{fig:boo1_figure}}    
\end{figure*}

The top row of Figure \ref{fig:boo1_figure} displays color-magnitude diagrams (CMDs) for the \booI\ dSph, which by virtue of its low declination is included in five of the six surveys considered here; for this reason, in what follows we will continue to use \booI\ as our example  for illustrating procedural details.  Red markers indicate stars projected within $1 R_{\rm half}$, which visibly trace the member sequence that is also consistent with the overplotted isochrone (age=10 Gyr, [Fe/H]$=-1.97$, distance modulus $\mu=19.67$). 

For each survey, Figure \ref{fig:boo1_figure} (second row) displays the observed density field of candidate members in \boo\ I.  In the figure, the 2D maps for the observed, best fit and residual stellar density fields have been rebinned from the original $250\times 250$ pixel$^2$ grid that we use in our analysis (see Section \ref{sec:spatialselection}) to a $50\times 50$ pixel$^2$ grid that we use for display purposes only.  The map of masked pixels retains the original $250\times 250$ pixel$^2$ grid.

\subsection{Spatial selection}
\label{sec:spatialselection}

For each galaxy and survey, we mask regions of poor data quality by first discretizing the square field of area $400 R_{\rm half}^2$ using a grid of $250\times 250$ square pixels (drawn in the tangent plane defined by the nominal galaxy center; see Section \ref{sec:analysis}), each having area $A_{\rm pix}\sim (0.08 R_{\rm half})^2$, ensuring that $\sim 500$ pixels resolve the region inside $1R_{\rm half}$.  We mask all pixels that fall outside the survey footprint.  We mask pixels falling within $3R_{\rm half}$ of the center of any other object (dwarf galaxies, star clusters and objects of ambiguous classification; here $R_{\rm half}$ is the projected halflight radius of the other object) that is included in the LVDB and has mean surface brightness $\mu_V\leq 25$ mag arcsec$^{-2}$ within its own halflight radius.  We mask pixels around bright stars listed in the USNO-B1.0 catalog \citep{usnob}; mask radius decreases with bright-star magnitude, from $\sim 7$ arcmin at magnitude $R=3$ mag to $\sim 0.1$ arcmin at $R=14$ mag.  In order to avoid crowding in central regions of the brightest galaxies, for all surveys except \textit{Gaia} (where we assume crowding is negligible) we mask pixels inside $0.1 R_{\rm half}$, $0.5R_{\rm half}$, $0.2R_{\rm half}$, and $0.1R_{\rm half}$ of Fornax, Leo I, Leo II and Sculptor, respectively. Figure \ref{fig:boo1_figure} (third row) displays the masks we use for \booI, with masked pixels colored black.  

For each survey and each galaxy, we estimate a crude `signal-to-noise' ratio that we compute as $S/N=(N_{\rm *,in}-N_{\rm *,out}A_{\rm in}/A_{\rm out})/\sqrt{N_{\rm *,in}}$, where $N_{\rm *,in}$ and $N_{\rm *,out}$ are the numbers of (unmasked, CMD-filtered) stars inside a circle of radius $2R_{\rm half}$ (twice the halflight radius listed in the LVDB) and outside a circle of radius $5R_{\rm half}$, respectively, and $A_{\rm in}$ and $A_{\rm out}$ are the areas over which those stars are selected.  We use $S/N$ as a rough proxy for sample quality.  For clarity in what follows, we show results derived only from catalogs with $S/N>4$ and, when multiple catalogs produce acceptable fits for a given galaxy, we present the result derived from the catalog giving the highest $S/N$ (unless we are explicitly comparing results from different catalogs, as in Figure \ref{fig:sbfit_survey_compare}).  In any case, results from all catalogs are available electronically (see Appendix).  

\section{Analysis}
\label{sec:analysis}

We use the distribution of observed stellar positions to model the stellar density field within a given dwarf galaxy. 

\subsection{Coordinates}
\label{sec:projection}
We use standard stereographic projection to map cataloged equatorial coordinates $(\alpha,\delta)$ onto position vectors $\vec{R}\equiv (\xi,\eta)$ in the plane that is tangent to the celestial sphere at the galaxy center $(\alpha_0,\delta_0)$, with \citep{snyder87} 
\begin{eqnarray}
\xi=k\cos(\delta)\sin(\alpha-\alpha_0),\hspace{1.15in}\\
\eta=k\bigl (\cos\delta_0\sin\delta-\sin\delta_0\cos\delta\cos(\alpha-\alpha_0)\bigr ),\nonumber
\label{eq:xi,eta}
\end{eqnarray}
and solid angle on sky related to area in the tangent plane by $d\Omega=k^{-2}d\xi d\eta$, where 
\begin{equation}
k=2\bigl(1+\sin\delta_0\sin\delta+\cos\delta_0\cos\delta\cos(\alpha-\alpha_0)\bigr )^{-1}
\label{eq:k}
\end{equation}
is the linear scale factor.  The stereographic projection is inverted by
\begin{eqnarray}
\delta=\sin^{-1}\biggl [\cos c\sin\delta_0+\frac{\eta\sin c\cos\delta_0}{R}\biggr ],\hspace{0.61in}\\
\alpha=\alpha_0+\tan^{-1}\biggl [\frac{\xi\sin c}{R\cos\delta_0\cos c-\eta\sin\delta_0\sin c}\biggr ],\nonumber
\label{eq:inverse_xi,eta}
\end{eqnarray}
where $R=\sqrt{\xi^2+\eta^2}$ and $c\equiv 2\tan^{-1}(R/2)$.

Compared with the gnomonic projection more typically adopted for mapping nearby dSphs, stereographic projection introduces  distortions that are isotropic instead of radially biased, and grow more slowly with angular separation from the tangent point.  At the extremities of our sample, where stars can be  $\approx 5^{\circ}-10^{\circ}$ from the centers of the largest galaxies, the linear distortion is $\approx 0.2\%-0.8\%$ for stereographic projection and $\approx 0.8\%-3\%$ (radial) and $\sim 0.4\%-1\%$ (tangential) for gnomonic projection. 

Simultaneously with the surface density profile discussed below, we infer the galaxy center by fitting its position $(\xi_0,\eta_0)$ in the nominal tangent plane that is defined by the nominal center listed in the LVDB.  For a given proposed center, we re-project all stars and grid points onto the corresponding tangent plane.  Our pixelized grid is regular only in the nominal tangent plane (where it is originally drawn), and has irregular spacing in planes corresponding to proposed centers at $(\xi_0,\eta_0)\neq (0,0)$.  In the recentered plane, the $i^{\rm th}$ pixel subtends solid angle
\begin{eqnarray}
\Omega_{\mathrm{pix},i}=\int_{\square_i} k^{-2}\bigl (\xi(p,q),\eta(p,q)\bigr ) \left | \frac{\partial(\xi,\eta)}{\partial (p, q)}\right | dp\,dq\nonumber\\
\approx k_i^{-2}\left | \frac{\partial (\xi,\eta)}{\partial (p,q)}\right |_i,
\label{eq:omega_pix}
\end{eqnarray}
where $(p,q)$ are continuous pixel index coordinates along the two grid axes and $\square_i$ signifies the unit cell corresponding to the $i^{\rm th}$ pixel.  In practice, we compute $\Omega_{\mathrm{pix},i}$ using the approximation given by the second line of Equation \ref{eq:omega_pix}, evaluating $k^{-2}\bigl (\xi(p,q),\eta(p,q)\bigr )$ and $\left | \partial(\xi,\eta)/\partial(p,q)\right |=\left |\frac{\partial\xi}{\partial p}\frac{\partial \eta}{\partial q}-\frac{\partial\xi}{\partial q}\frac{\partial\eta}{\partial p}\right |$ at the pixel center.  We use \texttt{numpy.gradient} to estimate the derivatives numerically as finite differences with unit spacing along both grid-index axes.  We then mask all pixels along the outer edges of the grid, where the gradients are estimated using one-sided differences in place of the central differences used for interior pixels.  

\subsection{Model}
\label{sec:bigsigma}
Following \citet{drlica-wagner20}, we assume that the number of stars, $N_{\rm obs}(\vec{R}),$ observed within a pixel of solid angle $\Omega_{\rm pix}(\vec{R})$ at position $\vec{R}$ is a Poisson random variable with expectation value $\Omega_{\rm pix}(\vec{R})\,\Sigma(\vec{R})$, where the surface number density $\Sigma(\vec{R})\equiv dN_{\rm obs}/d\Omega$ is approximately constant across a given pixel.  The Poisson model naturally accounts for the significant shot noise that is associated with star counts in the least luminous galaxies. 

We model the 2D stellar density field as the sum of contributions from a population of dwarf galaxy members and a population of nonmembers in the foreground: $\Sigma(\vec{R})=\Sigma_{\rm mem}(\vec{R})+\Sigma_{\rm non}(\vec{R})$, where $\Sigma_{\rm mem}\equiv dN_{\rm mem}/d\Omega$ and $\Sigma_{\rm non}\equiv dN_{\rm non}/d\Omega$.  We model the nonmember component as nearly uniform over the observed field.  However, because the fields of larger galaxies can stretch several degrees across the sky, we allow for a linear gradient such that 
\begin{equation}
\Sigma_{\rm non}(\vec{R})=\Sigma_{\rm non,0}\max\biggl (0,1+\vec{a}\cdot\frac{\vec{R}}{R_{\rm h,pub}}\biggr ).
\label{eq:bigsigma_non}
\end{equation}
where $\vec{a}=(a_{\xi}, a_{\eta})$ specifies the gradient and $R_{\rm h,pub}$ is the previously-published halflight radius from the LVDB.  

We model the member component as the elliptically-flattened projection of the spherically-symmetric 3D number density profile \citep{zhao96} 
\begin{equation}
\nu_{\rm mem}(r)=\nu_s\,\biggl(\frac{r}{r_s} \biggr)^{-\gamma}\biggl[1+\biggl(\frac{r}{r_s}\biggr)^{\alpha}\,\biggr]^{\frac{\gamma-\beta}{\alpha}}. 
\label{eq:abg}
\end{equation}
This  profile has index $d\log\nu_{\rm mem}/d\log r=-\beta$ at large ($r\gg r_s$) radius and $d\log\nu_{\rm mem}/d\log r=-\gamma$ at small ($r\ll r_s$), with $\alpha$ controlling the rate of transition.  It generalizes the standard \citet{plummer11} profile, which has $(\alpha,\beta,\gamma)=(2,5,0)$.  The standard Abel projection integral gives the 2D density of member stars in the tangent plane \citep{bt08}:
\begin{eqnarray}
    \tilde\Sigma_{\rm mem}(\vec{R})\equiv \frac{dN_{\rm mem}}{dA}=2\int_{R_e(\vec{R})}^{\infty}\frac{r\,\nu_{\rm mem}(r)\,dr}{\sqrt{r^2-[R_e(\vec{R})]^2}}\hspace{0.5in}\\
    =\Sigma_0X^{1-\gamma}\int_{0}^{\infty}(\cosh x)^{1-\gamma}\bigl[1+\bigl(X\cosh x\bigr)^{\alpha}\,\bigr]^{\frac{\gamma-\beta}{\alpha}}dx, \hspace{0.000025in}\nonumber
    \label{eq:bigsigma}
\end{eqnarray}
where $\Sigma_0=2r_s\nu_s$, $X\equiv R_e(\vec{R})/r_s$ and the second line follows from letting $\cosh x=r/R_e(\vec{R})$.  The `elliptical radius', $R_e(\vec{R})$, is the semi-major axis of the ellipse, having ellipticity $\epsilon\equiv 1-b/a$ and position angle $\theta$ ($\theta=0$ has the semi-major axis running north to south, $\theta=90^{\circ}$ has it east to west), that is centered on the origin in the tangent plane and intersects $\vec{R}$, satisfying \citep{martin08}
\begin{equation}
R_e^2(\vec{R})=
(\xi\sin\theta+\eta\cos\theta)^2+\frac{(\xi\cos\theta-\eta\sin\theta)^2}{(1-\epsilon)^2}.
\label{eq:r_elliptical}
\end{equation}
We assume ellipticity and position angle are global properties of a given galaxy, such that $R_e(\vec{R})$ is computed using the same values of $\epsilon$ and $\theta$ for all $\vec{R}$.  In practice, we evaluate the integral in Equation \ref{eq:bigsigma} numerically, obtaining $\log_{10}[\tilde \Sigma_{\rm mem}(X)/\Sigma_0]$ via multi-linear interpolation over a precomputed, regular grid that spans the four dimensions $(\alpha,\beta,\gamma,\log_{10}X)$.  Finally, we calculate the 2D density of member stars on the sky as
\begin{equation}
\Sigma_{\rm mem}(\vec{R})=k^2(\vec{R})\tilde\Sigma_{\rm mem}(\vec{R}).
\label{eq:on_sky}
\end{equation}

For comparison to the $\alpha\beta\gamma$ profile, we also consider for the member component the model of \citet{sersic68}, which specifies the 2D density (on the sky) profile directly as
\begin{equation}
\Sigma_{\rm mem,Sersic}(\vec{R})=\Sigma_0\exp\biggl [-\biggl (\frac{R_e(\vec{R})}{r_s}\biggr )^{1/n}\biggr ].
\label{eq:sersic}
\end{equation}
The \sersic\ model generalizes the exponential profile, which has $n=1$, that is routinely fit to dSph galaxies.  For smaller/larger $n$, the central concentration and inner curvature decrease/increase and the outer slope becomes steeper/shallower.  Thus while the \sersic\ model has a more flexible shape than either the Plummer or exponential models, its inner and outer slopes are coupled under a single value of $n$, allowing less flexibility than the $\alpha\beta\gamma$ model. 

We calculate scale densities $\Sigma_0$ and $\Sigma_{\rm non,0}$ in terms of free parameters $f_{\rm mem}$, the member fraction within the field, and $\log_{10}\bigl[N_{\rm field}/N_{\rm obs}\bigr]$, the ratio of model-predicted to observed numbers of stars within the field.  The predicted numbers of member and nonmember stars in the (unmasked) field are then $N_{\rm mem}^{\rm field}=f_{\rm mem}N_{\rm field}$ and $N_{\rm non}^{\rm field}=(1-f_{\rm mem})N_{\rm field}$, and the scale densities are $\Sigma_0=I_{\rm mem}^{-1}N_{\rm mem}^{\rm field}$ and $\Sigma_{\rm non,0}=I_{\rm non}^{-1}N_{\rm non}^{\rm field}$, where 
\begin{eqnarray}
I_{\rm mem}\equiv \frac{N_{\rm mem}^{\rm field}}{\Sigma_0}=\int_{\rm field}\frac{\Sigma_{\rm mem}(\vec{R})}{\Sigma_0}d\Omega,\hspace{0.25in}\label{eq:I}\\
I_{\rm non}\equiv \frac{N_{\rm non}^{\rm field}}{\Sigma_{\rm non,0}}= \int_{\rm field}\max\biggl (0,1+\vec{a}\cdot\frac{\vec{R}}{R_{\rm h,pub}}\biggr)d\Omega.\nonumber
\end{eqnarray}  
We use our $250\times 250$ pix$^2$ grid to calculate the scale densities numerically, approximating the integrals as $I_{\rm mem}\approx\sum_{i=1}^{N_{\rm pix}}\bigl(\Sigma_{\rm mem}(\vec{R}_i)/\Sigma_0\bigr)\Omega_{\rm pix}(\vec{R}_i)$ and $I_{\rm non}\approx \sum_{i=1}^{N_{\rm pix}}\max\bigl(0,1+\vec{a}\cdot \vec{R}/R_{\rm h,pub}\bigr)\Omega_{\rm pix}(\vec{R}_i)$, where $N_{\rm pix}$ is the number of unmasked pixels.
By construction, this parameterization enforces the constraint $N_{\rm field}=\int_{\rm field}\Sigma(\vec{R})d\Omega$.  

\subsection{Inference}
\label{sec:inference}
Given a vector of parameters $\vec{\Theta}$ that specifies the function $\Sigma(\vec{R})$, the $i^{\rm th}$ pixel contains $k_i$ stars with probability 
\begin{equation}
P(k_i|\vec{\Theta})=\frac{\exp\bigl [-\Omega_{\rm pix}(\vec{R}_i)\,\Sigma(\vec{R}_i)\bigr ]}{k_i!}\,\bigl (\Omega_{\rm pix}(\vec{R}_i)\,\Sigma(\vec{R}_i)\bigr )^{k_i},
\end{equation}
and the data set $D$ consisting of star counts in each of $N_{\rm pix}$ pixels (assuming these counts are uncorrelated) has likelihood $P(D\,|\,\vec{\Theta})=\prod_{i=1}^{N_{\rm pix}}P(k_i|\vec{\Theta})$.  In the continuum limit $\Omega_{\rm pix}\rightarrow 0$, and up to an additive constant that is model independent, the likelihood becomes that of the Poisson point process, with log-likelihood
\begin{equation}
\ln P(D\, |\,\vec{\Theta})=-N_{\rm field}+\sum_{i=1}^{N_{\rm obs}} \ln \Sigma(\vec{R}_i), 
\label{eq:loglike}
\end{equation}
where $N_{\rm field}\equiv\int_{\rm field}\Sigma(\vec{R})\,d\Omega$ is the predicted number of stars in the unmasked field and $N_{\rm obs}\equiv \sum_{i=1}^{N_{\rm pix}}k_i$ is the observed number.  While we use our pixelized grid to compute scale densities $\Sigma_0$ and $\Sigma_{\rm non,0}$ (see previous section), we evaluate the log-likelihood in Equation \ref{eq:loglike} using discrete stellar positions.  

Our model for $\Sigma(\vec{R})$ has $12$ free parameters in the case of the $\alpha\beta\gamma$ model, $10$ free parameters in the case of the \sersic\ model.  Table \ref{tab:parameters} identifies each parameter, gives a brief description, and specifies the ranges of uniform prior probability distribution functions (PDFs) we adopt; we use the same priors for every galaxy and every survey.  

\begin{deluxetable*}{llllllllllllll}
\tablewidth{0pt}
\tablecaption{Free parameters of models for stellar density field.
\label{tab:parameters}
}
\tablehead{\colhead{parameter}&\colhead{description}&\colhead{uniform prior range}}
\startdata
$\xi_0/\mathrm{arcmin}$&$\xi$-component of offset from published centroid &$[-3,+3]$\\
$\eta_0/\mathrm{arcmin}$&$\eta$-component of offset from published centroid &$[-3,+3]$\\
$\log_{10}\bigl [\frac{N_{\rm field}}{N_{\rm obs}}\bigr ]$&ratio of predicted to observed number of stars in the unmasked field (Eq. \ref{eq:loglike})&$[-1,+1]$\\
$f_{\rm mem}$&member fraction among observed stars&$[0,+1]$\\
$\log_{10}[r_s/R_{\rm h,pub}]$&scale radius (Eq. \ref{eq:abg})&$[-1,+1]$\\
$\alpha$&sharpness of power-law break (Eq. \ref{eq:abg})&$[0.5,3]$\\
$\beta$&outer power-law index (Eq. \ref{eq:abg})&$[3.1,10]$\\
$\gamma$&inner power-law index (Eq. \ref{eq:abg})&$[-1,2]$\\
$n$&\sersic\ index (Eq. \ref{eq:sersic}) &$[0.5,10]$\\
$\epsilon\equiv 1-b/a$&ellipticity (Eq. \ref{eq:r_elliptical})&$[0,1]$\\
$\theta$&position angle (Eq. \ref{eq:r_elliptical})&$[-90^{\circ},+90^{\circ}]$\\
$a_{\xi}/R_{\rm h,pub }^{-1}$&$\xi$-component of foreground density gradient&$[-0.01,+0.01]$\\
$a_{\eta}/R_{\rm h,pub}^{-1}$&$\eta$-component of foreground density gradient&$[-0.01,+0.01]$\\
\enddata
\end{deluxetable*}

We use Bayesian methods to infer the posterior PDF of model parameters, given data set $D$:
\begin{equation}
P(\vec{\Theta} | D)=\frac{P(D | \vec{\Theta})\pi(\vec{\Theta})}{P(D)},
\end{equation}
where $P(D | \vec{\Theta})$ is the likelihood of the data, given the model (Eq. \ref{eq:loglike}), $\pi(\vec{\Theta})$ is the prior PDF for model parameters (Table \ref{tab:parameters}; the adopted prior is separable into the product of 1D PDFs for each parameter), and $P(D)=\int_{\vec{\Theta}} P(D | \vec{\Theta})\pi(\vec{\Theta})d\vec{\Theta}$ is the marginal likelihood, or `evidence'.  We use the nested sampling \citep{skilling04} algorithm MultiNest \citep{feroz09}, implemented in Python as PyMultiNest \citep{buchner14}, to estimate the evidence and obtain random samples from the posterior PDF.  

For each galaxy and survey, we fit eight different models: the full $\alpha\beta\gamma$ model with parameters and priors specified in Table \ref{tab:parameters},  the full \sersic\ model as specified in Table \ref{tab:parameters}, a \citet{plummer11} model with $(\alpha,\beta,\gamma)$ held fixed at $(2,5,0)$, an exponential model that fixes the \sersic\ index to $n=1$, and spherically-symmetric versions of all four, with ellipticity and position angle held fixed at $\epsilon=\theta=0$.  The fits of Plummer and exponential models enable comparisons to many previously-published results and, via the computed value of the Bayesian evidence, provide a quantitative means for model selection.  The fits of spherical models may be relevant for subsequent analyses that assume spherical symmetry, as is the case in many standard kinematic models (e.g., \citealt{read21} and references therein).  

After each fit, we bin the data according to elliptical radius (using ellipticity and position angle from the best-fitting model), and compute the statistic $\chi^2\equiv \sum_{i=1}^{N_{\rm bin}}\bigl (\Sigma_{\rm obs}\bigl(R_e(\vec{R}_i))-\Sigma_{\rm fit}(R_e(\vec{R}_i))\bigr)^2/\sigma^2_{\Sigma_{\rm obs}(R_e(\vec{R}_i))}$, where $\Sigma(R_e(\vec{R}_i))$ is the best-fitting model, $\Sigma_{\rm obs}(R_{\rm e}(\vec{R}_i))=N_{\mathrm{obs},i}/A_i$ is the number of stars counted in the $i^{\rm th}$ elliptical annulus, divided by the annulus area, and $\sigma_{\Sigma_{\rm obs}(R_e(\vec{R}_i))}=\sqrt{N_{\mathrm{obs},i}}/A_i$ is the Poisson noise.  

In what follows, we show results derived only from catalogs that have $S/N\geq 4$, and fits for which $0.5\leq \chi^2/N_{\rm bin}\leq 1.5$.

\subsection{Implied Parameters}
We use the samples from posterior PDFs to obtain corresponding posteriors for the halflight radius, number of member stars, and total stellar mass.  The inferred number of members that are within adopted magnitude limits and have elliptical radius smaller than $R_e$ satisfies
\begin{equation}
\frac{N_{\rm mem}^{\rm obs}(R_e)}{N_{\rm mem}^{\rm obs}(\infty)}=\frac{\displaystyle\int_0^{R_e/r_s}X\Sigma_{\rm mem}(X)dX}{\displaystyle\int_0^{\infty}X\Sigma_{\rm mem}(X)dX},
\label{eq:n_re}
\end{equation}
where 
\begin{equation}
N_{\rm mem}^{\rm obs}(\infty)=2\pi r_s^2(1-\epsilon)\int_0^{\infty} X\Sigma_{\rm mem}(X)dX
\label{eq:n_obs}
\end{equation}
is the total number of members within adopted magnitude limits (including any stars within masked regions or outside the adopted field).  For the $\alpha\beta\gamma$ model, we compute numerically the semimajor axis, $a_{\rm half}$, of the ellipse containing half of the member stars.  The 2D halflight `radius' is then $R_{\rm half}=a_{\rm half}\sqrt{1-\epsilon}$.  We report $R_{\rm half}$ in physical units of pc, propagating uncertainty in $a_{\rm half}$, ellipticity and the host dSph's distance modulus, with the last sampled from a Gaussian distribution with mean and standard deviation as listed in the LVDB; in the three cases where the LVDB-listed distance modulus has asymmetric errorbars, we sample a Gaussian with standard deviation set to half the difference between upper and lower bounds.  

For the \sersic\ model, we use the analytic results
\begin{equation}
a_{\rm half,Sersic}=r_sb_n^n,
\end{equation}
and 
\begin{equation}
N_{\rm mem,Sersic}^{\rm obs}(\infty)=2\pi r_s^2\Sigma_0(1-\epsilon)n\Gamma(2n),
\end{equation}
with $b_n$ satisfying $2\int_0^{b_n}x^{2n-1}e^{-x}dx=\Gamma(2n)$ and approximated according to \citet[][their Equation 18]{ciotti99}. 

We infer the total number of member stars, correcting for photometric incompleteness, by combining information from the adopted isochrone with an adopted mass function, $\phi(M)\equiv dN/dM$.  Given age,  metallicity and distance, the isochrone relates stellar mass to luminosity and (with bolometric corrections) filter magnitude, letting us translate the adopted faint and bright magnitude limits, $m_{\rm faint}$ and $m_{\rm bright}$, respectively, into mass limits $M_{\rm faint}$ and $M_{\rm bright}$.  We take into account galaxy distance errors when translating apparent magnitude limits to stellar mass limits: for each random sample from the posterior, we assign a dwarf galaxy distance modulus by randomly drawing from a Gaussian distribution with mean and standard deviation given by the central value and uncertainty listed in the LVDB.  

Assuming the stellar mass function is normalized so that $\int_{M_{\rm min}}^{M_{\rm max}}\phi(M)dM=1$, we calculate the expected total number of member stars, corrected for photometric incompleteness, as 
\begin{equation}
N_{\rm mem}^{\rm tot}=N_{\rm mem}^{\rm obs}(\infty)\frac{1}{\displaystyle\int_{M_{\rm faint}}^{M_{\rm bright}}\phi(M)dM},
\label{eq:ntot}
\end{equation}
where $M_{\rm min}=0.1M_{\odot}$ represents approximately the hydrogen-burning limit of the main sequence, and $M_{\rm max}$ (computed by \texttt{Minimint} based on MIST isochrones) is the maximum mass of stars surviving to the adopted age.  If $M_{\rm bright}>M_{\rm max}$, we replace the integration upper limit $M_{\rm bright}$ with $M_{\rm max}$ in denominator.  Similarly, we compute expected values for the total stellar mass and total $V$-band luminosity as
\begin{eqnarray}
M_{\rm mem}^{\rm tot}=N_{\rm mem}^{\rm tot}\displaystyle\int_{M_{\rm min}}^{M_{\rm max}}\phi(M)MdM;\hspace{0.3in}\nonumber\\
L_{\rm V,mem}^{\rm tot}=N_{\rm mem}^{\rm tot}\displaystyle\int_{M_{\rm min}}^{M_{\rm max}}\phi(M)L_{V}(M)dM,\hspace{0.3in}
\label{eq:m_l_tot}
\end{eqnarray}
where $L_V(M)$ is the interpolated $V$-band luminosity corresponding to mass $M$.  

Compared to alternative procedures that directly sample the mass function and apply photometric selection effects until matching the observed number of stars (e.g., \citealt{martin08}), ours neglects shot noise and its dependence on magnitude, which can become significant when estimating the luminosities of the faintest galaxies \citep{martin08,martin16}.  However, shot noise has relatively little effect on the estimated stellar mass $M_{\rm mem}^{\rm tot}$, which for this reason is a more useful diagnostic of stellar content \citep{smith24}.  Our analysis also neglects uncertainties in age/metallicity that set the relationship between mass and filter magnitudes; however, \citet{martin08} show that these uncertainties are negligible up to $\sim 0.5$ dex in metallicity and $\pm 2$ Gyr in age, and tend to be dominated by galaxy distance errors.

We consider the stellar mass functions formulated by \citet{salpeter55}, \citet{kroupa02} and \citet{chabrier03}.  The Salpeter function is a power law with $\phi(M)\propto M^{-2.35}$.  The Kroupa function that we adopt is a broken power law, with $\phi(M)\propto M^{-2.3}$ for masses $M/M_{\odot}\geq 0.5$ and $\phi(M)\propto M^{-1.3}$ for $0.1\leq M/M_{\odot}<0.5$.  The Chabrier function is log-normal, with mean $\log_{10}[M/M_{\odot}]=0.08$ and standard deviation $\sigma_{\log_{10}[M/M_{\odot}]}=0.7$.  

\subsection{Validation}
\label{sec:validation}
In order to gauge reliability, we apply our fitting procedure to mock catalog data that we generate to mimic the real catalogs in terms of sample size and survey footprint.  The basic framework of modeling the dSph stellar density field using unbinned catalog data is already well established \citep[e.g.,][]{martin08,bechtol15,drlica-wagner20}, so we focus specifically on our inferences of shape parameters.  For a given real galaxy, we generate ten mock data sets that differ according to the input values of $\beta$ and $\gamma$ (we regard the transition parameter $\alpha$ as a nuisance parameter to be marginalized over).  Holding the inner (outer) index fixed at the Plummer value of $\gamma=0$ ($\beta=5$), we generate mock data having a member component with the ellipticity and position angle set to the published LVDB-listed values, with input values of the outer (inner) index of $\beta=3.5,5.0,6.5,8.0,9.5$ ($\gamma=0.0,0.5,1.0,1.5,2.0$), with input transition parameter $\alpha=2$, and a foreground component having uniform density.  We draw the appropriate numbers of stars to ensure that, after adopting the same pixel masks applied to the real catalog data (using the galaxy-specific catalogs listed in Table \ref{tab:sbfit_table}), the mock data sets contain the same numbers of stars and the same member fractions within the field that we infer from the real data when fitting the $\alpha\beta\gamma$ model.  For each mock data set, we repeat our fitting procedure for both $\alpha\beta\gamma$ and Plummer models.  

\begin{figure*}
    \begin{tabular}{@{}llll@{}}
    \renewcommand{\arraystretch}{0}
    \includegraphics[width=0.6\textwidth,trim=0.2in 0.1in 0in 0in, clip]{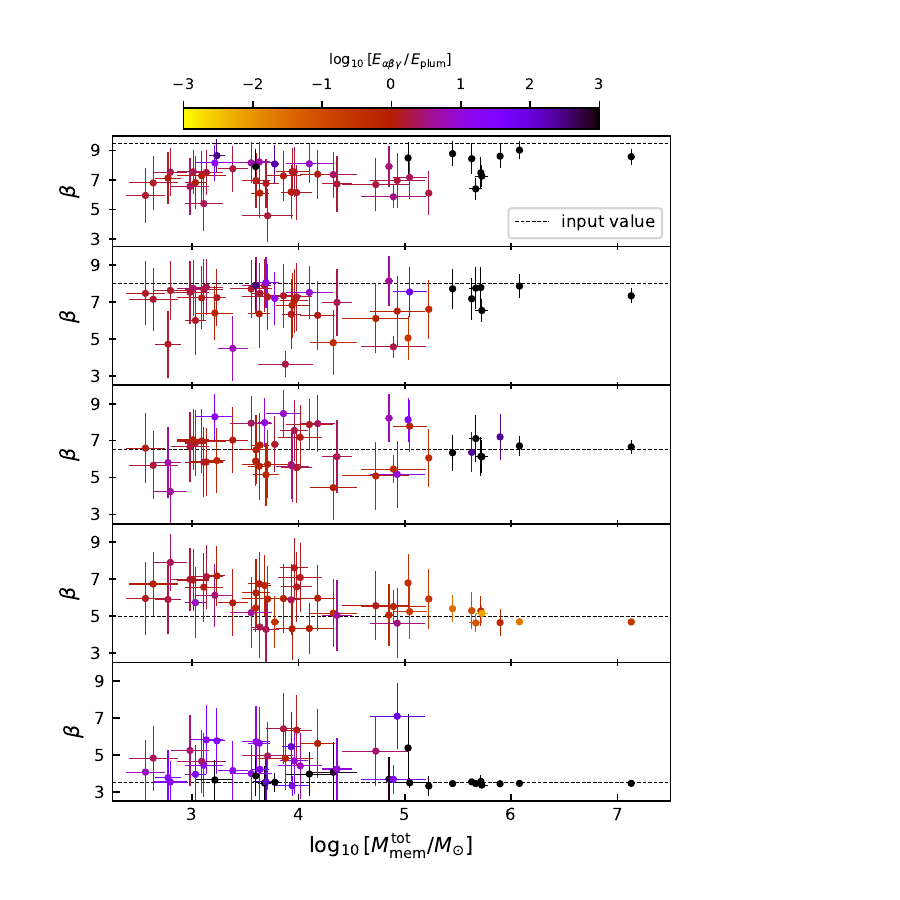
    }& \hspace{-1in}\includegraphics[width=0.6\textwidth,trim=0.2in 0.1in 0in 0in, clip]{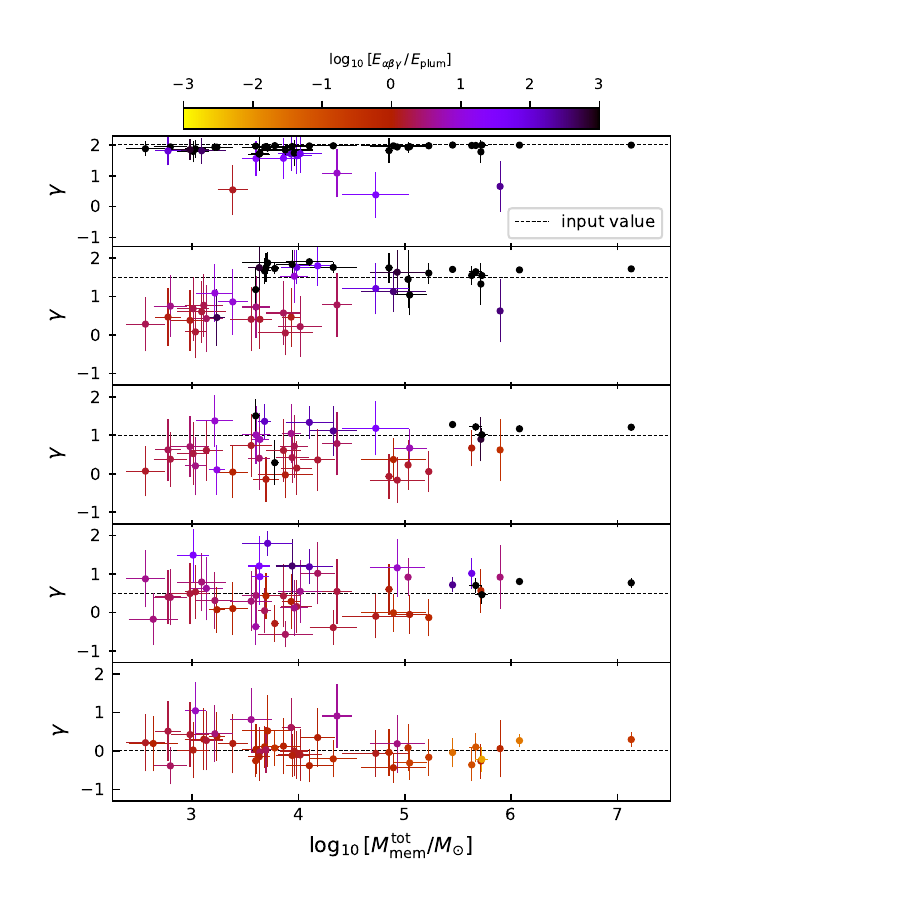}\\
    
    \end{tabular}
    \caption{Recovery of input outer (left) and inner (right) power-law indices $\beta$ and $\gamma$, respectively, vs. inferred stellar mass, respectively, from fits of the $\alpha\beta\gamma$ model to mock data generated to mimic---in terms of area and depth---the real survey (listed in Table \ref{tab:sbfit_table}) data available for each observed dwarf galaxy.  Horizontal lines indicate input values; realizations shown in the left (right) panel all have Plummer-like $\gamma=0$ ($\beta=5$) and $\alpha=2$ input values.  Marker color indicates (logarithm of) the evidence ratio that aides model selection, with positive values favoring the $\alpha\beta\gamma$ model over the Plummer model.   }
\label{fig:check_systematics}
\end{figure*}

Figure \ref{fig:check_systematics} displays, as a function of the stellar mass inferred from the $\alpha\beta\gamma$ fit to the real catalog data, the values of $\beta$ (left panels) and $\gamma$ (right panels) that we infer for each mock data set (error bars indicate 68\% credibility intervals).  The observed behavior is reassuring, as the distributions of inferred $\beta$, $\gamma$ shift according to input values.  Most input values are recovered within errorbars, especially when the posterior is not dominated by priors; for reference, when summarized in terms of mean and standard deviation, our uniform priors are $\beta=6.5\pm 2.0$, $\gamma=0.5\pm 0.9$.  Posterior intervals tend to be similar to these priors at $M_{\rm mem}^{\rm tot}\lesssim10^4M_{\odot}$, indicating that the available data for the lowest-mass dSphs are uninformative regarding profile shape.  At higher masses, where the shape parameters are relatively well constrained, evidence ratios tend to favor the correct input model.  These results give confidence that we can reliably infer $\beta$ and $\gamma$ for the more massive dSphs, with sufficient sensitivity to detect and characterize departures from the standard Plummer profile with many of the present samples. 

\section{Results}
\label{sec:results}

We provide results from our fits of all models to all galaxies in all surveys at the Zenodo database, https://doi.org/10.5281/zenodo.20600393.  Products include input catalogs of color/magnitude-and spatially-selected sources, pixel masks, random samples from posterior PDFs for model parameters and implied parameters (e.g., $R_{\rm half}$, $N_{\rm mem}^{\rm tot}$, $M_{\rm mem}^{\rm tot}$, $L_{\rm V,mem}^{\rm tot}$) for the $\alpha\beta\gamma$, \sersic, Plummer and exponential models, both flattened and spherical cases.  We also provide electronic tables that include summary statistics (mean, standard deviation, quantiles) of all fitted and derived parameters for all galaxies, surveys and models.  The Appendix provides details of the information contained in the electronic data files.   

For more immediate convenience, Table \ref{tab:sbfit_table} summarizes posterior PDFs from our fits of the elliptical $\alpha\beta\gamma$ model to individual galaxies having at least one catalog meeting the criteria stated at the end of Section \ref{sec:inference}.  The second column identifies the catalog producing the highest S/N for that galaxy.  Listed quantities then indicate means and standard deviations from MultiNest's sampling of posterior inferred from the fit to that catalog, including central coordinates, $\alpha$, $\beta$\, $\gamma$, scale radius, ellipticity, position angle, number of observed members $N_{\rm mem}^{\rm field}=f_{\rm mem}N_{\rm field}$, (circularized) halflight radius, and total stellar mass.  

Returning to the example of \booI, the fourth row of Figure \ref{fig:boo1_figure} displays best-fitting (elliptical) $\alpha\beta\gamma$ models for each survey.  The fifth row shows normalized residuals between observation and best-fitting $\alpha\beta\gamma$ model, $\bigl(\Sigma_{\rm obs}(\vec{R})-\Sigma_{\rm \alpha\beta\gamma}(\vec{R})\bigr)/\sigma_{\Sigma_{\rm obs}(\vec{R})}$, where $\sigma_{\Sigma_{\rm obs}(\vec{R})}=\sqrt{N_{\rm obs}(\vec{R})}\,\Omega^{-1}_{\rm pix}(\vec{R})$ is the Poisson noise.  The residual fields usually have the appearance of noise. 
The bottom row of Figure \ref{fig:boo1_figure} shows how the stellar density inferred for \booI\ under the $\alpha\beta\gamma$ model compares to the stellar density observed in bins of elliptical radius, $\Sigma_{\rm obs}(R_e(\vec{R}))$ (calculated using the ellipticity of the best-fitting $\alpha\beta\gamma$ model); we reiterate that our fits are to the positions of individually-cataloged stars and not to the binned data.  Gray, red and blue envelopes represent the central 68\% credible intervals inferred for $\Sigma(R_e(\vec{R}))$, $\Sigma_{\rm mem}(R_e(\vec{R}))$ and $\Sigma_{\rm non}(R_e(\vec{R}))$, respectively.  

\subsection{Model vs. Model}
\label{sec:modelvmodel}
\begin{figure}
\includegraphics[width=3.5in,trim=0.1in 1.65in 2.5in 0.25in, clip]{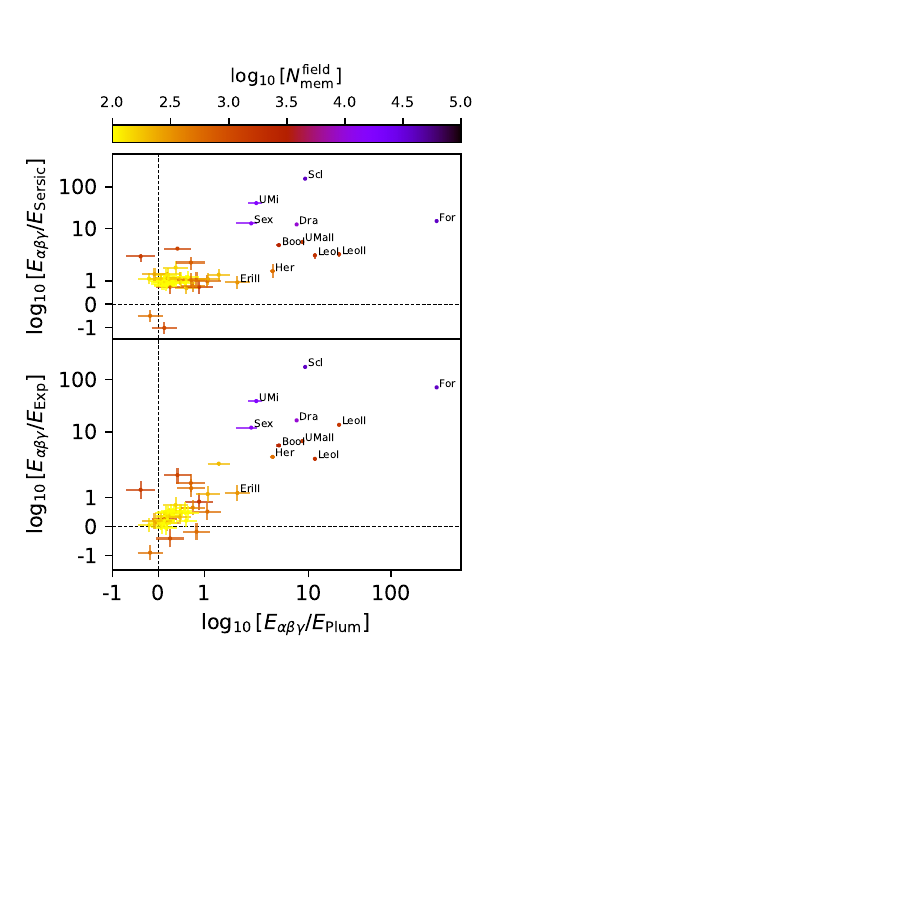}
    \caption{Ratios of the marginalized likelihood, or `evidence', comparing the $\alpha\beta\gamma$ model to Plummer, \sersic\ and exponential models, for the galaxies and surveys listed in Table \ref{tab:sbfit_table}.  Color indicates the inferred number of observed member stars.  Where the ratio is decisive ($\bigl|\log_{10}[E_{\alpha\beta\gamma}/E_{\rm other}]\bigr|>1.5$), it favors the $\alpha\beta\gamma$ model over the Plummer, \sersic\ and exponential models; text identifies dSphs meeting this criterion.    \label{fig:dlogevidence}}    
\end{figure}

Figure \ref{fig:dlogevidence} displays evidence ratios comparing the elliptical $\alpha\beta\gamma$, \sersic, Plummer and exponential models, with color indicating the inferred number of observed member stars, $N_{\rm mem}^{\rm field}$.  Where the evidence ratio is decisive, with $\bigl |\log_{10}[E_1/E_2]\bigr |>1.5$, we typically have $N_{\rm mem}^{\rm field}\gtrsim10^3$, identifying a target sample size for  model selection.  For our samples, this threshold roughly corresponds to a stellar mass of $M_{\rm mem}^{\rm tot}\gtrsim10^4M_{\odot}$.  Indeed, our tests with mock data (Section \ref{sec:validation}, Figure \ref{fig:check_systematics}) show that above this threshold in mass, model selection tends to be decisive and shape parameters are reliably constrained.  

In all cases where the evidence ratio is decisive, the $\alpha\beta\gamma$ model is preferred over the Plummer, \sersic\ and exponential models.  Evidently, among these options, dSph density fields are best characterized by the flexible double power law.  For this reason we do not consider the \sersic\ and exponential models further, and use the Plummer model only to demonstrate potential systematic errors associated with model selection.

\begin{figure*}
\includegraphics[width=6.5in,trim=0.25in 4.in 0.45in 0.4in, clip]{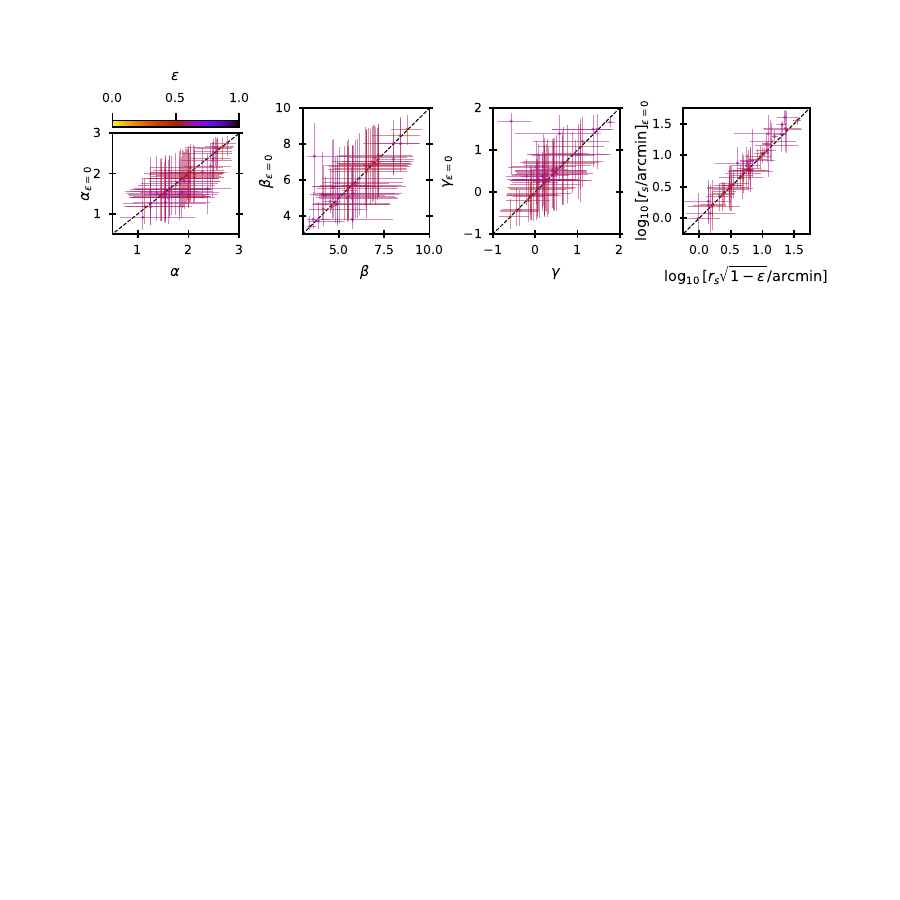}
    \caption{Comparisons of transition parameter $\alpha$ (left), outer power-law index $\beta$ (middle-left), inner power-law index $\gamma$ (middle-right), and scale radius $r_s$ (right, adjusted to the geometric mean using the inferred ellipticity), inferred under the circular (vertical axis) versus elliptical (horizontal axis) $\alpha\beta\gamma$ model.  Color indicates (central value of) the inferred ellipticity.  \label{fig:sbfit_check_systematics_sph}}    
\end{figure*}

For the $\alpha\beta\gamma$ model, Figure \ref{fig:sbfit_check_systematics_sph} compares values of $\alpha$, $\beta$, $\gamma$ and the scale radius, $r_s$, inferred for the galaxies/surveys listed in Table \ref{tab:sbfit_table} under our models that allow for elliptical morphology and those that enforce circular symmetry ($\epsilon=0$).  We find excellent agreement, implying that our inferences of shape and scale parameters are insensitive to the choice between circular and elliptical models.  Furthermore, the right-most panel of Figure \ref{fig:sbfit_check_systematics_sph} demonstrates that the standard practice of `circularizing' the elliptical scale radius via the $\sqrt{1-\epsilon}$ multiplier gives a good representation of the scale radius inferred under the circular model.  

\subsection{Survey vs. Survey}
\label{sec:surveyvsurvey}

\begin{figure*}
    \begin{tabular}{@{}llll@{}}
    \renewcommand{\arraystretch}{0}
    \includegraphics[width=3.5in,trim=0.25in 0.25in 2.0in 0.25in, clip]{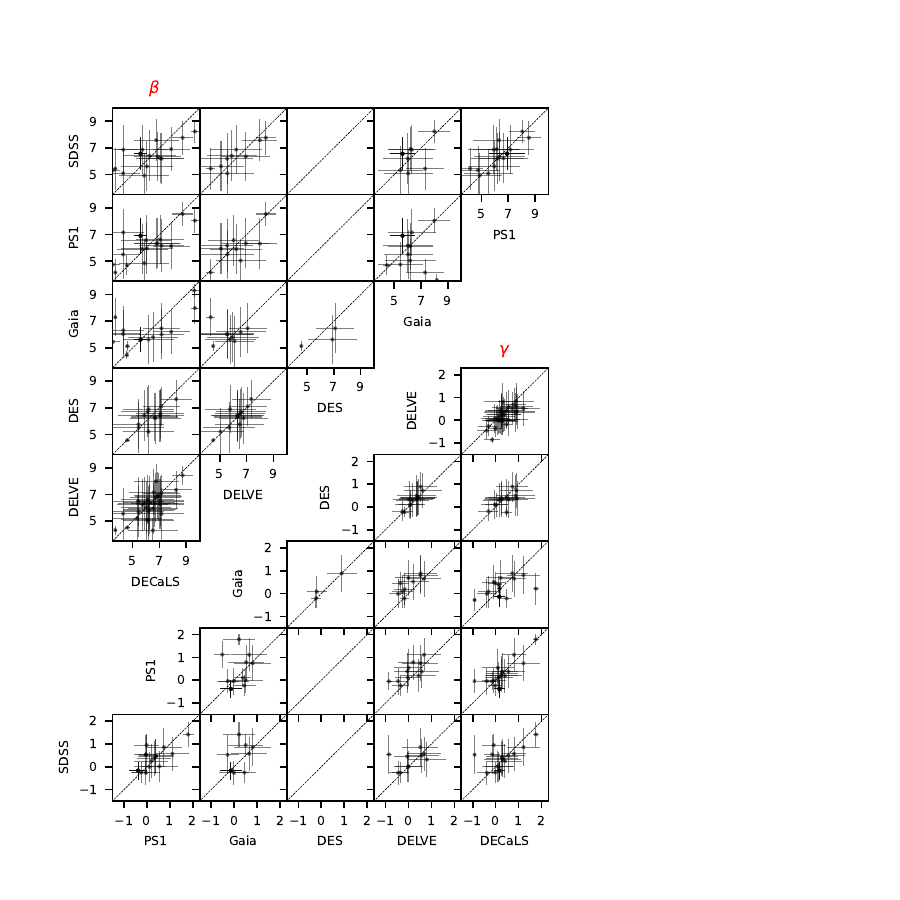}& \hspace{-0in}\includegraphics[width=3.5in,trim=0.25in 0.25in 2.0in 0.25in, clip]{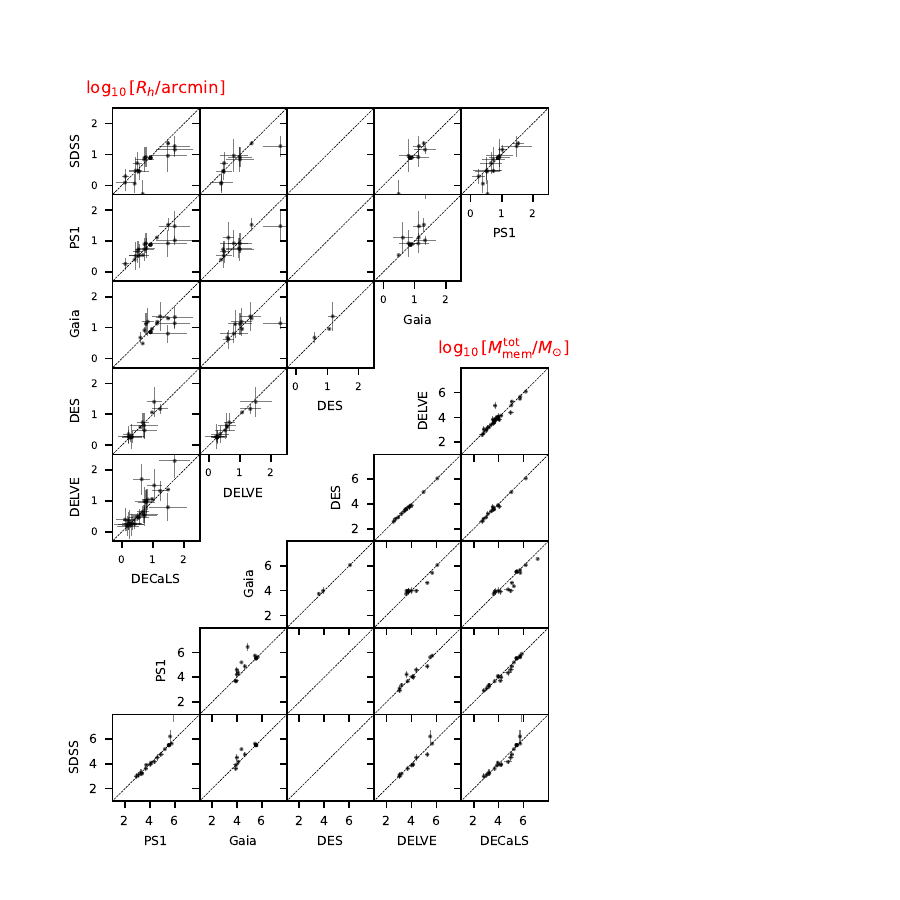}
    \end{tabular}
    \caption{Comparisons of values inferred for outer power-law index $\beta$ (top left), inner power-law index $\gamma$ (bottom left), projected halflight radius $R_{\rm half}$ (top right), and total stellar mass $M_{\rm mem}^{\rm tot}$ (bottom right) from different survey catalogs for Milky Way satellites with acceptable fits multiple surveys.\label{fig:sbfit_survey_compare}}
\end{figure*}

Figure \ref{fig:sbfit_survey_compare} compares values of $\beta$, $\gamma$, $R_{\rm half}$ and $M_{\rm mem}^{\rm tot}$ inferred for individual galaxies using data from different survey catalogs.  We reiterate that the DES, DELVE and DECaLS catalogs are not entirely independent; however, SDSS, PS1, \textit{Gaia} are all wholly independent of each other and of DES, DELVE and DECaLS.  The observed correlations in Figure \ref{fig:sbfit_survey_compare}---even among the fully independent surveys---give confidence that our constraints on these parameters are repeatable and convey real information about the internal structure of the observed galaxies.

\subsection{Covariance}
\begin{figure*}
\includegraphics[width=6.5in,trim=0in 0in 0in 0.in, clip]{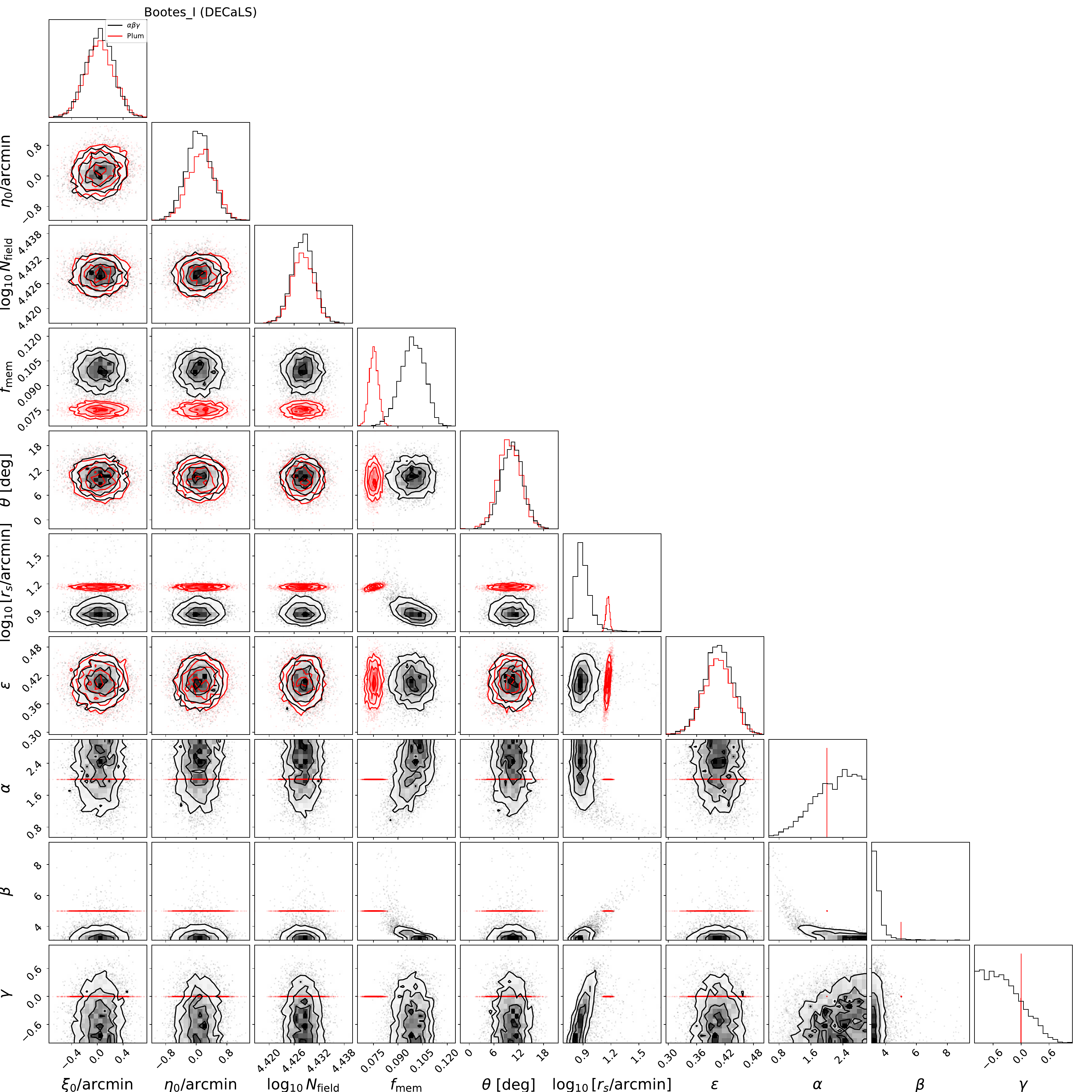}
\caption{Covariance in the posterior probability distribution among free parameters in the $\alpha\beta\gamma$ (black) and Plummer (red) stellar density models, shown here for our fits to \booI/DECaLS data.  \label{fig:Bootes_I_decals_sb_corner1_abg}}    
\end{figure*}

Again using \booI\ as an example, the `corner' plot in Figure \ref{fig:Bootes_I_decals_sb_corner1_abg} shows covariance among ten individual model parameters (for legibility we omit the two parameters controlling the foreground density gradient).  We observe correlations among the power-law indices $\beta$, $\gamma$, and the scaling parameters $r_s$ and $f_{\rm mem}$ (which sets the scale density $\Sigma_0$).  Both $\beta$ and $\gamma$ exhibit positive correlations with the scale radius.  The outer index $\beta$ anti-correlates with the member fraction, with shallower steeper outer slopes corresponding to lower fractions.    

\begin{figure}
\includegraphics[width=3.5in,trim=0.in 0in 0in 0in, clip]{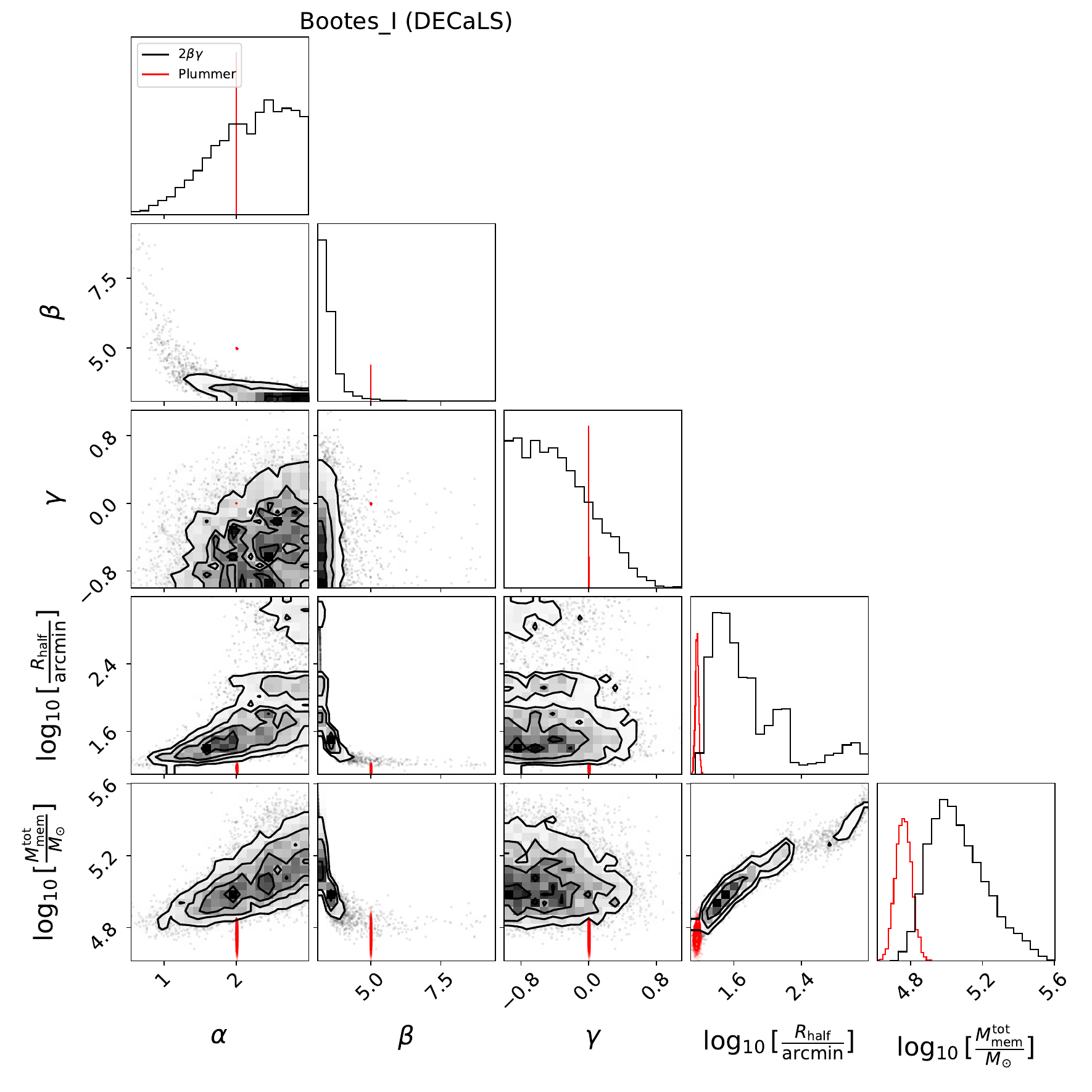}
    \caption{Covariance in posterior probability distribution among $\alpha$, $\beta$, $\gamma$ and the implied (projected) halflight radius and total stellar mass, shown here for our fits of $\alpha\beta\gamma$ (black) and Plummer (red) models  to \booI/DECaLS data.    \label{fig:Bootes_I_decals_sb_corner2_abg}}    
\end{figure}

Figure \ref{fig:Bootes_I_decals_sb_corner2_abg} shows how these correlations with $\beta$ and $\gamma$ map onto standard structural parameters like the halflight radius and stellar mass, $M_{\rm mem}^{\rm tot}$ (for simplicity, estimated here using only the Kroupa mass function).  We find that when the outer density slope is shallow ($\beta<5$), the index $\beta$ exhibits a strong anti-correlation with $R_{\rm half}$ and, given an intrinsic correlation between $R_{\rm half}$ and $M_{\rm mem}^{\rm tot}$, a strong anti-correlation with $M_{\rm mem}^{\rm tot}$.  This behavior follows from the fact that the total number of member stars, and hence both $R_{\rm half}$ and $M_{\rm mem}^{\rm tot}$, become infinite when $\beta\leq 3$.  While our adopted prior range $3.1\leq \beta\leq 10$ excludes this extremity, if the data allow slowly-declining outer density profiles ($\beta \lesssim4$), then the halflight radius and total stellar mass can be much larger than would be inferred under standard models that assume a fixed profile shape.  

\begin{figure*}
    \includegraphics[width=6.5in,trim=0.in 0.75in 0in 1.in, clip]{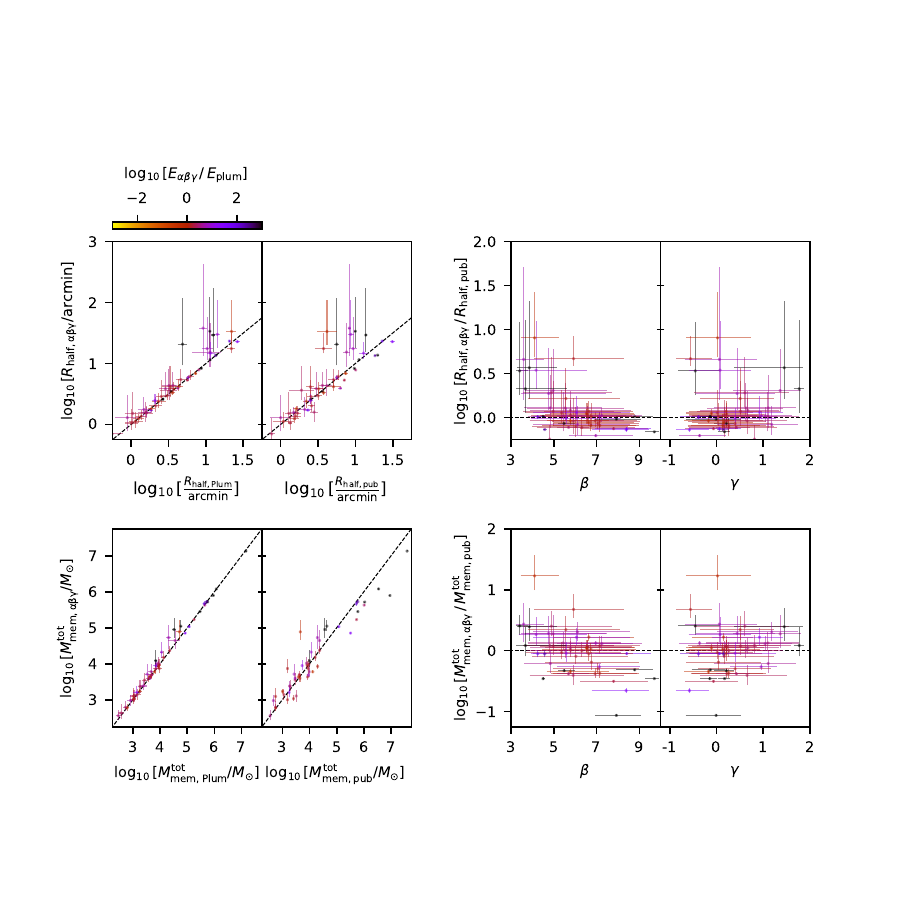}
    \caption{Comparison of (projected) halflight radii (top) and stellar masses (bottom) inferred under the $\alpha\beta\gamma$ model to those inferred from the standard Plummer model and to  previously-published results from the Local Volume Database.  Far-left and middle-left panels compare estimates of $R_{\rm half}$ and $M_{\rm mem}^{\rm tot}$ from the $\alpha\beta\gamma$ model directly to those from the Plummer model and to published results, respectively; middle-right and right panels show how ratios of structural parameters estimated under the $\alpha\beta\gamma$ model and published values vary with the outer and inner logarithmic slopes, $\beta$ and $\gamma$, that we infer.  \label{fig:sbfit_check_systematics_rhalf2d_mstar_evidence}}
\end{figure*}

Making this point explicitly, the left-most panels of Figure \ref{fig:sbfit_check_systematics_rhalf2d_mstar_evidence} directly compare estimates of halflight radii (top) and stellar masses (bottom) that we obtain when fitting the $\alpha\beta\gamma$ and Plummer models.  The structural parameters estimated under the $\alpha\beta\gamma$ model tend to be larger and more uncertain than those estimated under the Plummer model, in some cases with halflight radii and stellar masses reaching values up to an order of magnitude larger under the $\alpha\beta\gamma$ model.  Not surprisingly, given the Plummer model's popularity, we find similar behavior when comparing $\alpha\beta\gamma$ results to previously-published halflight radii and stellar masses (middle-left panels)---except that the stellar masses we infer under the $\alpha\beta\gamma$ model are systematically \textit{smaller} than published values for the most massive dSphs ($M_{\rm mem}^{\rm tot}\gtrsim10^5M_{\odot}$; bottom row, middle-left panel).  The reasons for this discrepancy are unclear, but likely include 1) heterogeneous methodology among previous efforts to extrapolate star counts to total stellar masses (e.g., different assumptions about the stellar mass function, which can be difficult to discern for previously-published estimates) and 2) the observed anti-correlation of total stellar mass with $\beta$, coupled with the fact that we infer steep ($\beta>7$) outer profiles for several of the high-mass dSphs (see Section \ref{sec:diversity}).  Indeed, the right two columns of panels in Figure \ref{fig:sbfit_check_systematics_rhalf2d_mstar_evidence} show that ratios of $\alpha\beta\gamma$-inferred to published values of both halflight radius and stellar mass correlate especially with $\beta$, further illustrating the importance of measuring rather than adopting profile slopes. 

Even when the halflight radii and masses inferred under $\alpha\beta\gamma$ and Plummer models are consistent, as they are for most galaxies, the posterior PDFs for the $\alpha\beta\gamma$ model are significantly broader.  This fact reflects the $\alpha\beta\gamma$ model's broader prior, which fully contains the Plummer prior.  In Table \ref{tab:sbfit_table}, the dagger ($\dagger)$ symbol marks entries for which, when stated in terms of the associated linear quantity, the standard deviation is larger than the median (e.g., Segue 1's halflight radius of $\log_{10}[R_{\rm half}/\mathrm{pc}]=$\seglogabg\ corresponds to $R_{\rm half}=$ \seglinabg\ pc).  In most of these cases, the corresponding estimate under the Plummer model is `resolved' (e.g., Segue 1 has $\log_{10}[R_{\rm half}/\mathrm{pc}]=$\seglogplum, corresponding to $R_{\rm half}=$ \seglinplum\ pc under the Plummer model, in good agreement with the LVDB value).  It is important to note, however, that the tighter constraints obtained under the Plummer model are driven entirely by the prior.  Unless there is justification for the narrower prior, observational constraints on halflight radii and stellar masses of many dSphs are significantly weaker than has been apparent under Plummer and/or exponential models that impose a fixed profile shape.  This result underscores the recent finding by \citet{andersson25}, whose cosmological simulations resolve extended stellar halos among the stellar populations of faint dSph analogs, leading them to conclude that halflight radii and stellar masses are prone to underestimation when inferred using simple density models with fixed shape.  

\subsection{Centroids}

\begin{figure}
\includegraphics[width=3.in,trim=0.25in 2.5in 2.75in 0in, clip]{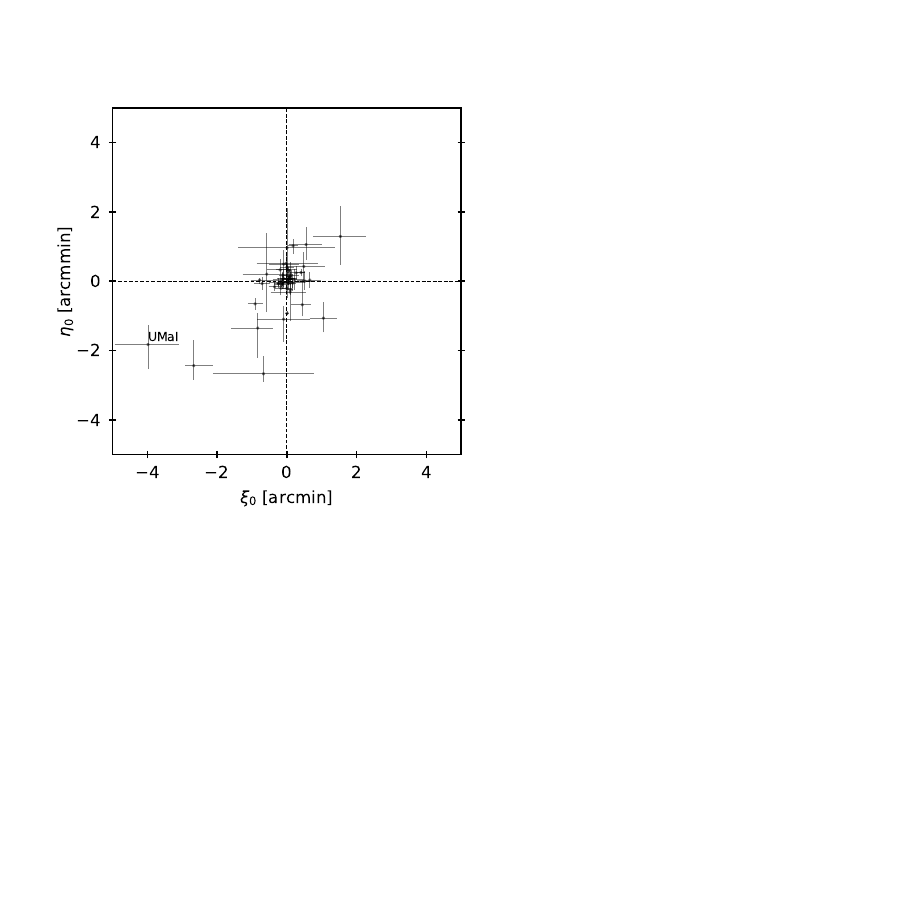}
    \caption{Position of the galaxy centroid, in the tangent plane with origin at the previously-published centroid, for the galaxies/surveys listed in Table \ref{tab:sbfit_table}.  \label{fig:sbfit_check_systematics_centroid1}}    
\end{figure}

Figure \ref{fig:sbfit_check_systematics_centroid1} shows offsets of the centroids we infer, under the $\alpha\beta\gamma$ model, from the nominal centroids listed in the LVDB.  In most cases we find good agreement to within measurement uncertainties, which is reassuring since many of the nominal centroids come from discovery papers that predate the availability of Gaia-based astrometric calibration.  

The most notable outlier is Ursa Major I (UMaI), for which we infer a centroid $(\alpha_0,\delta_0)=(158.6729 \pm 0.0158,+051.9244 \pm 0.0089)$ that is $\approx 4$ arcmin west-southwest\footnote{Upon noticing that posteriors from our initial fits to UMaI's centroid were skewed toward the western edge of our uniform prior for $(\xi_0,\eta_0)$, we expanded the prior range for both components to $\pm 6$ arcmin.} of the nominal value  $(158.7706, +51.9479)$.  The LVDB value comes from deep CFHT/Megacam imaging by \citet{munoz18}, who discard one of their two UMaI fields due to poor weather conditions.  Their remaining field skews northeast of UMaI's center (their Figure 21), likely affecting the inferred centroid.  The centroid we infer stands in much better agreement with earlier measurements of $(158.703\pm 0.012,+51.93\pm 0.32)$ by \citet[][based on SDSS photometry]{martin08} and $(158.6850,+51.9261)$ by \citet[][Subaru/Suprime-Cam photometry]{okamoto08}.

\subsection{Diversity of Structural Parameters}
\label{sec:diversity}
\begin{figure}
    \includegraphics[width=3.5in,trim=0.25in 0.25in 2.3in 0.5in, clip]{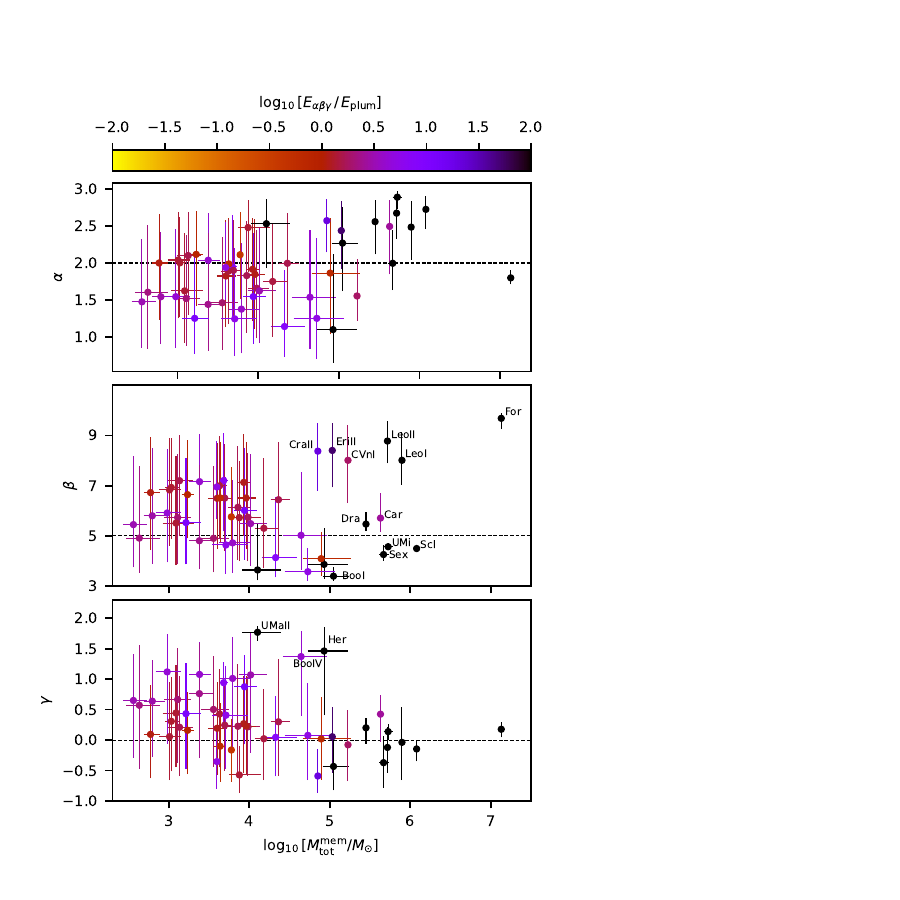}
    \caption{Shape parameters $\alpha$,  $\beta$ and $\gamma$ vs. stellar mass, inferred for \ngood\ Milky Way satellites.  Dashed lines indicate Plummer values $(\beta,\gamma)=(5,0)$.  Marker color indicates (logarithm of) the evidence ratio that aides model selection, with positive values favoring the $\alpha\beta\gamma$ model over the Plummer model.  \label{fig:sbfit_beta_gamma}}
\end{figure}

Figure \ref{fig:sbfit_beta_gamma} displays the inferred values of $\alpha$, $\beta$ and $\gamma$ as functions of the inferred stellar mass (using the Kroupa mass function), with marker color indicating the evidence ratio, $\log_{10}[E_{\alpha\beta\gamma}/E_{\rm Plum}]$.  As expected from our tests with mock data (Section \ref{sec:validation}), it is for masses $M_{\rm mem}^{\rm tot}\lesssim10^4 M_{\odot}$ that the evidence ratio tends to be inconclusive for model selection and our inferences about shape parameters tend to be dominated by priors.  At higher masses, the evidence ratio tends to favor the $\alpha\beta\gamma$ model decisively, with $\log_{10}[E_{\alpha\beta\gamma}/E_{\rm Plum}]>1.5$ for \nabg\ galaxies: \abgs.

\begin{figure*}
    \includegraphics[width=6.5in,trim=0.25in 1.6in 0.53in 0.25in, clip]{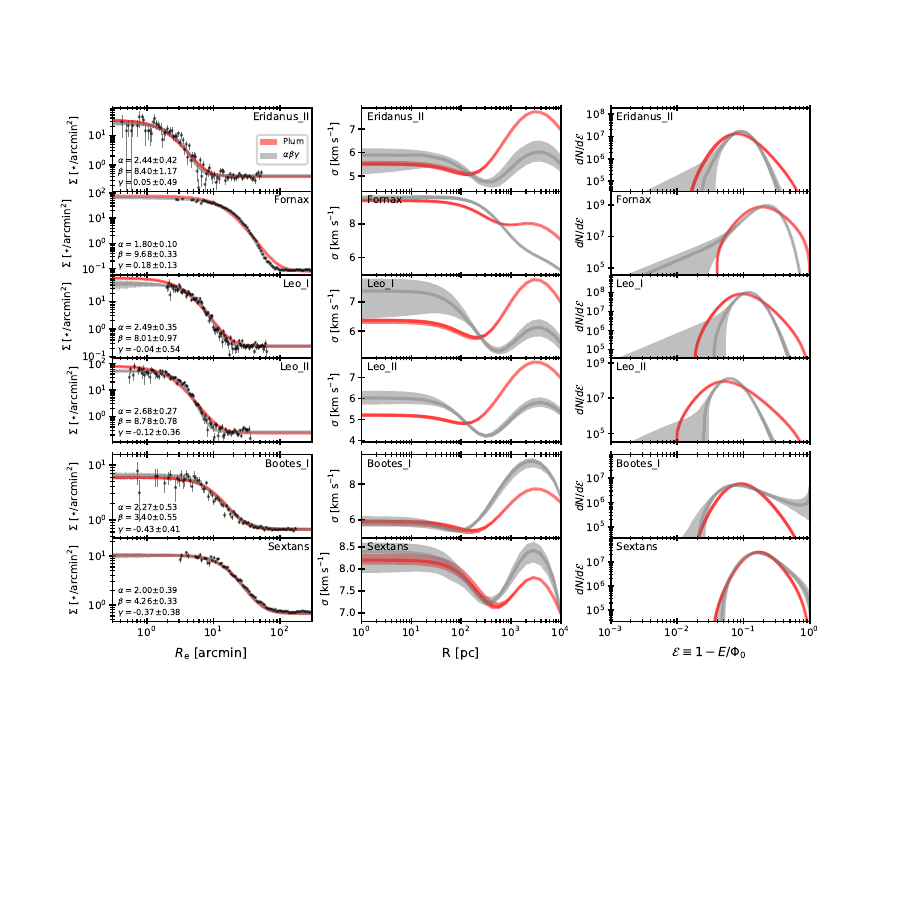}
    \caption{\textit{Left:} stellar surface density profiles for dwarf galaxies inferred to have outer slopes that are steeper ($\beta>5$; top four panels) than a Plummer profile, and shallower ($\beta<5$; bottom two panels) than a Plummer profile, with 68\% credible intervals overplotted for the fitted $\alpha\beta\gamma$ (gray) and Plummer (red) models.  \textit{Center:} corresponding velocity dispersion profiles calculated from the Jeans equation, assuming spherical symmetry, velocity dispersion isotropy, and a gravitationally-dominant dark matter halo that follows the \citet{navarro97} form with mass $M_{200}=10^9M_{\odot}$ and concentration $c_{200}=10$.  \textit{Right:} corresponding stellar-orbital energy distribution within the fiducial NFW dark matter halo.
\label{fig:special_beta}}
\end{figure*}

For these relatively massive dSphs, we find a variety of interesting constraints.  First, Eridanus II (\betaEridanusII) Leo I (\betaLeoI), Leo II (\betaLeoII) and Fornax (\betaFornax) all exhibit outer density profiles that decline more steeply than standard Plummer profiles.  In contrast, the outer profiles of \booI\ (\betaBootesI) and Sextans (\betaSextans) decline more gently (lower left).  The left-hand panels of Figure \ref{fig:special_beta} display observed stellar density profiles for each of these galaxies, with $\alpha\beta\gamma$ and Plummer profiles overlaid for comparison.  

In the inner regions, \booI\ (\gammaBootesI), Carina (\gammaCarina), Draco (\gammaDraco), Eridanus II (\gammaEridanusII), Fornax (\gammaFornax), Leo I (\gammaLeoI), Leo II (\gammaLeoII), Sculptor (\gammaSculptor), Sextans (\gammaSextans) and Ursa Minor (\gammaUrsaMinor) all are consistent with Plummer-like cores of uniform density.   

\begin{figure*}
    \includegraphics[width=6.5in,trim=0.25in 3.35in 0.53in 0.25in, clip]{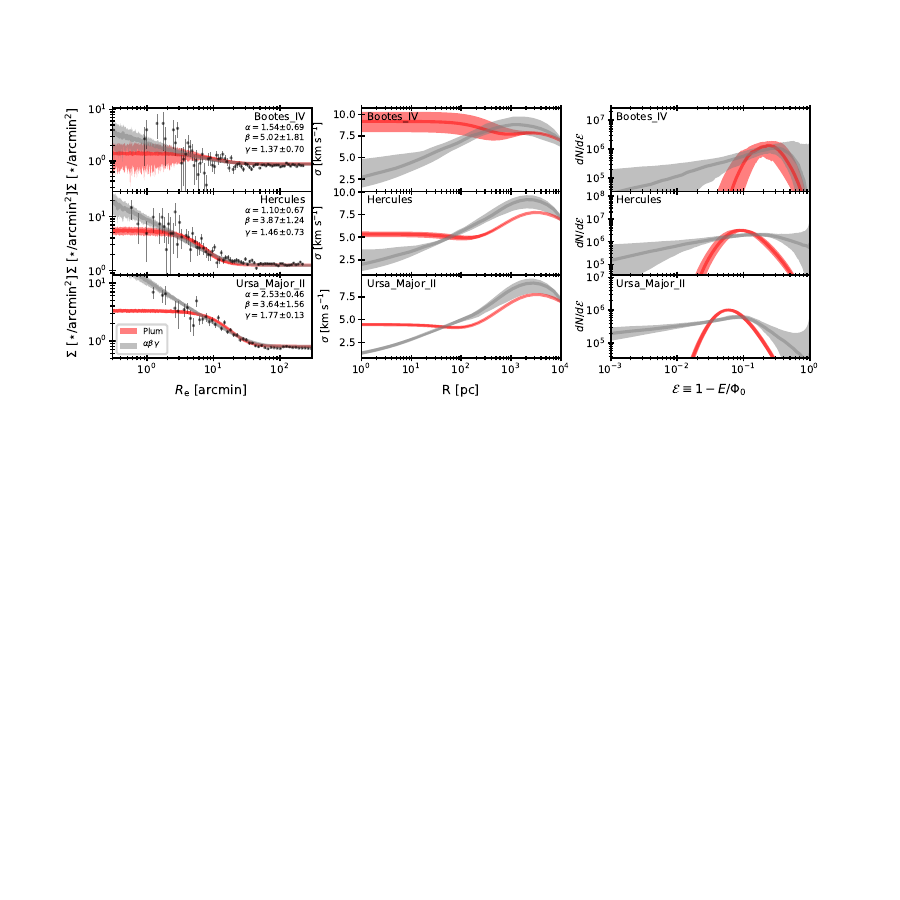}
    \caption{Similar to Figure \ref{fig:special_beta}, but for systems inferred to have stellar cusps ($\gamma>0$).  
\label{fig:special_gamma}}
\end{figure*}

Perhaps most surprisingly, we identify two galaxies at intermediate stellar mass ($4\lesssim\log_{10}[M_{\rm mem}^{\rm tot}/M_{\odot}]\leq 5$) that prefer steep stellar cusps:  Hercules (\gammaHercules) and Ursa Major II (\gammaUrsaMajorII).  A third, \boo\ IV (\gammaBootesIV), also prefers a stellar cusp, but the evidence for the $\alpha\beta\gamma$ model is less persuasive, with $\log_{10}[E_{\alpha\beta\gamma}/E_{\rm Plum}]\approx 0.86$.
The left-hand panels of Figure \ref{fig:special_gamma} show that, even to the eye, the steeply cusped profiles fit the observed inner density profiles of these galaxies better than the Plummer profile.  

Thus, at least for the relatively well measured galaxies with $M_{\rm mem}^{\rm tot}\gtrsim10^4M_{\odot}$, we observe a diversity of stellar structure, inferring centrally cusped as well as cored density profiles, and outer profiles that decline more steeply, and others that decline more gently, than the standard Plummer profile.

\subsection{Dynamical Implications of Model Selection}
The shapes of the stellar density profiles necessarily have implications for dynamical mass modeling.  In order to gauge significance of the observed structural diversity of Milky Way dSphs in this context, we compare line-of-sight velocity dispersion profiles corresponding to our fits of (spherical, with $\epsilon=0$) $\alpha\beta\gamma$ and Plummer models.  For simplicity, we compute these profiles assuming isotropy of the velocity dispersion tensor, in which case the spherically-symmetric Jeans equation for the line-of-sight velocity dispersion profile simplifies to \citep{mamon05}
\begin{equation}
\sigma_{\rm los}^2(R)=\frac{2G}{\Sigma_{\rm mem}(R)}\displaystyle\int_R^{\infty}\sqrt{1-\frac{R^2}{r^2}}\nu(r)M(r)\frac{dr}{r},
\label{eq:mamon}
\end{equation}
where $M(r)$ is the mass, including any contribution from dark matter, enclosed within a sphere of radius $r$.  Again for simplicity, we assume the mass profile is dominated by a dark matter halo following the form proposed by \citet{navarro97},
\begin{equation}
M(r)=M_{200}\,g_c\biggl(\ln\bigl(1+r/r'_s\bigr)-\frac{r/r'_s}{1+r/r'_s}\biggr),
\end{equation}
where $g_c\equiv \bigl(\ln(1+c_{200})-c_{200}/(1+c_{200})\bigr)^{-1}$
and we choose mass $M_{200}=10^9M_{\odot}$ and concentration $c_{200}=10$ to represent plausible values for halos hosting dSphs.  The corresponding velocity dispersion profiles appear in the center-column panels of Figures \ref{fig:special_beta} and \ref{fig:special_gamma}.  We see in Figure \ref{fig:special_beta} that when the $\alpha\beta\gamma$ model has outer stellar density falling more steeply (gently) than Plummer as in the top four (bottom two) panels, use of the $\alpha\beta\gamma$ model predicts a velocity dispersion profile that is systematically smaller (larger) than the one that would be predicted using the Plummer model---especially in the outer regions, where the Plummer fits to the stellar density profiles are systematically above (below) the $\alpha\beta\gamma$ fits.  

Figure \ref{fig:special_gamma} shows more complicated systematic differences when the $\alpha\beta\gamma$ fit favors a cusp, with use of the $\alpha\beta\gamma$ model predicting velocity dispersions smaller than would be predicted under the Plummer model in inner regions, and larger at intermediate radii---again reflecting the fact that the Plummer fits to the stellar density profiles are systematically below the $\alpha\beta\gamma$ fits in inner regions, and systematically above at intermediate radii.  

The observed discrepancies between profiles calculated from $\alpha\beta\gamma$ and Plummer models can reach a few km s$^{-1}$, potentially exceeding random observational errors.  Thus given the diversity of profile shapes we infer for the known dSphs, the use of models sufficiently flexible to capture that diversity is crucial for accurate mass modeling.  In a companion paper by \citet{splawska26}, we examine the effect of model selection on dynamical mass estimators of the form $M(r_0)=\mu r_0\langle\sigma^2\rangle/G$, where $r_0$ is the radius of the sphere enclosing mass $M(r_0)$, $\langle\sigma^2\rangle$ is the surface-brightness-weighted mean (squared) velocity dispersion, and coefficient $\mu$ depends on both $\nu(r)$ and $M(r)$.

For the same models with velocity dispersion profiles shown in the center columns of Figures \ref{fig:special_beta} - \ref{fig:special_gamma}, rightmost columns show how the relative orbital energy, $\mathcal{E}\equiv 1-E/\Phi_0$, where $E\equiv \frac{1}{2}v^2+\Phi$ and $\Phi$ is the gravitational potential with central value $\Phi_0$, is distributed among stars.  We use the technique of \citet{eddington16} to compute the phase-space distribution function $f(\mathcal{E})$, which we then sample using the procedure described by \citet{errani20}.\footnote{Our application of Eddington's method reveals some of the posterior volume---particularly where $\gamma$ approaches zero---to be physically incompatible with the adopted NFW halo, with $f(\mathcal{E})<0$ for some $\mathcal{E}$ (see also, e.g., \citealt{sanchez-almeida23}).  As our intention here is merely to illustrate potential systematic errors due to choice of stellar density model, and not to infer parameters of the dark matter halo, we do not attempt to remove such models.}  In Figure \ref{fig:special_beta}, we see that, as $\mathcal{E}$ becomes large and stars are more weakly bound, dSphs with steep (shallow) outer slopes $\beta>5$ ($\beta<5$) have truncated (extended) energy distributions.  Toward the other extreme, as $\mathcal{E}\rightarrow 0$ and stars become tightly bound, Figure \ref{fig:special_gamma} shows that dSphs with central stellar cusps ($\gamma>1$) populate the most bound energy states with relatively high probability, making them especially resilient to disruption by external tidal forces \citep{errani20,errani24}.  The best-fitting Plummer models miss this attribute altogether, again highlighting the importance of model flexibility.  

\section{Comparison to Previous Work}
\label{sec:previous}

We are not the first to measure structural parameters across the Milky Way dSph population using homogeneous methodology.  Previous work by, e.g.,  \citet{ih95}, \citet{martin08}, \citet{okamoto12} and  \citet{munoz18} establish the genre.  The novelty here comes from our application of the relatively flexible $\alpha\beta\gamma$ model to data from several recent and large-scale sky surveys.  The aforementioned studies have considered various fitting functions, but none with freedom to vary inner and outer density profiles independently.  Our work establishes that this capability is important, enabling superior fits even after penalizing, as the evidence calculation does, for the extra degree of freedom.  

Despite differences in data and methodology, many of our measurements evoke previous results qualitatively.  Steeply falling profiles even beyond the `break' radii where King models fail have previously been noted in Fornax and Leo I \citep{battaglia06,wang19,moskowitz20,sohn07}, foreshadowing our result that $\beta>7$ for these galaxies.  We do not find previous discussion of steep outer profiles in Eridanus II or Leo II specifically; however, King and \sersic\ fits by \citet[][their Figure 10]{munoz18} to deep CFHT/Megacam photometry are able to fit Leo II's outer density profile only by systematically overshooting the observed inner density.  Unlike the \sersic\ model, the $\alpha\beta\gamma$ model can reconcile a cored inner profile (\gammaLeoII) in Leo II with a steep outer profile (\betaLeoII). 

We detect shallow outer profiles in Sextans (\betaSextans) and especially \boo\ I (\betaBootesI).  Analyzing DECam photometry for \boo\ I, \citet{roderick16} find that the King (resp. exponential) model gives acceptable fits to the inner (outer) density profile but not to the outer (inner) profile.  To the eye, the outer profile that they observe (their Figure 7) appears to follow a shallow power law, which would be consistent with our measurement.  Similarly, the \boo\ I density profile measured from CFHT/Megacam photometry by \citet{munoz18} appears to decline more gently in the outer regions than the best-fitting King and \sersic\ models overplotted in their Figure 13.  

In their fits of two-component exponential models to 60 dSphs, \citet{jensen24} identify \boo\ I as the dSph with the most prominent extended outer stellar component, which they consider in the context of accretion and/or tidal disruption events.  This result tracks with our finding that \boo\ I exhibits the shallowest outer slope of any dSph.  Of the remaining eight dSphs for which \citet{jensen24} detect a second exponential component, we obtain meaningful constraints on the outer slopes only for Sculptor (\betaSculptor) and Ursa Minor (\betaUrsaMinor).  Interestingly, these are the two dSphs we find to be most Plummer-like in terms of $\beta$ and $\gamma$.  In fact, both galaxies overwhelmingly favor the standard Plummer model over the exponential (Figure \ref{fig:dlogevidence}), with $\log_{10}[E_{\rm Plum}/E_{\rm exp}]>30$.  This result is entirely compatible with Jensen et al.'s detection in both galaxies of outer stellar profiles that decay more slowly than the single exponential model. 

For Sextans, \citet{roderick16b} fit King, Plummer and exponential models to density profiles derived from DECam photometry, finding the best fit using the Plummer model.  \citet{cicuendez18} obtain good fits using all three, as well as for the \sersic\ model, and \citet{munoz18} show good fits of King and \sersic\ models to CFHT/Megacam-derived Sextans profiles.  These three previous studies give little reason to anticipate our measurement of a shallow outer profile in Sextans.  Yet we consistently recover a shallow outer profile whether we use DECaLS (\sextansbetadecals), DELVE (\sextansbetadelve), or PS1 (\sextansbetapsone) catalogs for Sextans.  Using SDSS (\sextansbetasdss; the SDSS footprint truncates within the Sextans field) and/or \textit{Gaia} (\sextansbetagaia), Sextans' outer slope is relatively poorly constrained.   

As for the inner density profiles, our most statistically significant detections of stellar cusps occur for Hercules (\gammaHercules) and Ursa Major II (\gammaUrsaMajorII), for both of which the $\alpha\beta\gamma$ model is overwhelmingly favored over the Plummer model (Figure \ref{fig:dlogevidence}).  \citet{sand09} fit King, Plummer, and exponential models to deep Large Binocular Telescope photometry of Hercules.  Sand et al. conclude that all three models show excellent agreement with the observed density profiles, but inspection of their Figure 5 reveals that only the exponential model, which alone among the models used by Sand et al. has a (mild) cusp when deprojected to 3D \citep{vitral20}, has sufficiently high density to fit the innermost bin. 

\citet{munoz18} do not comment explicitly on the inner density profile of Ursa Major II, derived from their deep CFHT/Megacam photometry, but their Figure 8 clearly shows a large central density that is poorly fit by the overplotted King and \sersic\ models.  In earlier analysis, presumably based on the same observations, \citet{munoz10} state that the Ursa Major II data are best fit by a double power law with inner index $d\log\Sigma/d\log R=-0.96$ and outer index $d\log\Sigma/d\log R=-2.40$ which, after reducing each index by one to account crudely for projection, matches remarkably well our measurements of \gammaUrsaMajorII\ and \betaUrsaMajorII.

Based on their analysis of \textit{Gaia} data, \citet[][`V26' hereafter]{vitral26} report a \textit{deficit} of central Fornax stars, with an inner density profile that decreases toward the center ($\gamma<0$).  Comparison to our results for Fornax, \gammaFornax\ for DECaLS and, for more direct comparison,  \gaiagammaFornax\ specifically for \textit{Gaia}, is complicated by the fact that V26 fit a 2D $\alpha\beta\gamma$ model directly to the projected stellar density field, whereas we fit the projection of the 3D $\alpha\beta\gamma$ model.  While the $\alpha\beta\gamma$ shape parameters are not equivalent across 2D and 3D implementations, in both cases $\gamma<0$ would imply a central deficit of stars.  For our fits to the Fornax catalogs derived from DECaLS and \textit{Gaia}, respectively, our fits have $9\%$ and $11\%$ of the posterior probability at $\gamma<0$.  Thus even using our projection of the 3D $\alpha\beta\gamma$ model, a central deficit of Fornax stars cannot be ruled out.\footnote{For the most direct comparison with V26's result, we also fit their 2D $\alpha\beta\gamma$ model to our Fornax/$\textit{Gaia}$ sample (which extends over a smaller field than V26's sample), again with the uniform prior over $-1\leq\gamma_{\rm 2D}\leq +2$.  In this case we obtain $\gamma_{\rm 2D}=0.02\pm 0.06$, peaking near a perfect core, but with $\approx 40\%$ of the posterior probability at $\gamma_{\rm 2D}<0.$  The evidence ratio indicates no preference between the 3D and 2D $\alpha\beta\gamma$ models.} 

Finally, our results do not directly contradict those of \citet{sanchez-almeida24}, who find that the stacked \textit{HST} surface density profiles of six ultrafaint dSphs are best fit by a cored model.  Two (Hydra II, Sagittarius II) of the six of the ultrafaint dSphs in their study are too faint to produce acceptable fits according to our quality criteria (Section \ref{sec:analysis}).  For the remaining four (Horologium I, Horologium II, Phoenix II, Triangulum II), the ground-based survey data that we use are more spatially complete, but significantly shallower than the photometry derived from central \textit{HST} pointings analyzed by \citet{sanchez-almeida24}.  However, the structural diversity that we find across the dSph population undermines justification for fitting multiple dSphs with a single profile.

\section{Summary \& Discussion}
\label{sec:discussion}
We have fit a flexible double power-law $\alpha\beta\gamma$ stellar density model to catalog-level public survey data for the Milky Way's \ngal\ known dSph satellites, obtaining satisfactory fits for \ngood\ dSphs.  For comparison we have fit standard Plummer, \sersic\ and exponential models as well, considering elliptical and circularly-symmetric versions in all cases.  

We have learned the following:
\begin{itemize}

\item In all cases where the evidence ratio is decisive---typically where the number of observed member stars is $N_{\rm mem}^{\rm field}\gtrsim10^3$, and/or the inferred stellar mass is $M_{\rm mem}^{\rm tot}> 10^4M_{\odot}$)---the $\alpha\beta\gamma$ model is favored over the Plummer, \sersic\ and exponential models (Figure \ref{fig:dlogevidence}). 

\item We infer a wide range of $(\beta,\gamma)$ across the dSph population, indicating diversity of structure among dSph stellar components (Table \ref{tab:sbfit_table} and Figure \ref{fig:sbfit_beta_gamma}).  Among the most massive dSphs, Eridanus II, Fornax, Leo I and Leo II all have outer density profiles that decline more steeply ($\beta\gtrsim8$) than Plummer profiles.  The inner profiles of dSphs with stellar mass $\gtrsim10^5M_{\odot}$ have shallow slopes that are consistent with cores ($\gamma\approx 0$) of uniform stellar density.  At lower mass the inner slopes are poorly constrained, except in a few ultrafaint dSphs (e.g., Hercules, Ursa Major II) where we infer steep stellar cusps ($\gamma\gtrsim1.5$).  These results are broadly consistent across different surveys (Figure \ref{fig:sbfit_survey_compare}).

\item Estimates of structural parameters (halflight radius, total stellar mass) tend to be both larger and more uncertain under the $\alpha\beta\gamma$ model than under the special case of the Plummer model (Figure \ref{fig:sbfit_check_systematics_rhalf2d_mstar_evidence}).  In cases where the outer density profile decays slowly ($\beta\lesssim4$), structural parameters can be up to an order of magnitude larger under $\alpha\beta\gamma$ than under the Plummer model.  

\item The previous item is reflected also when comparing to previously-published results (Figure \ref{fig:sbfit_check_systematics_rhalf2d_mstar_evidence}).  That is, structural parameters estimated under the $\alpha\beta\gamma$ model are more uncertain than published values, and up to an order of magnitude larger when the inferred outer slope is shallow.  However, we also find that stellar masses inferred under $\alpha\beta\gamma$ are significantly \textit{smaller} than published values for dSphs with stellar mass $\gtrsim10^5M_{\odot}$, the regime where several dSphs have steep outer slopes ($\beta\gtrsim8$).

\item Inferences of the stellar density profile shape parameters are robust to whether the model allows for (globally) elliptical morphology or assumes spherical symmetry (Figure \ref{fig:sbfit_check_systematics_sph}).
\item
The use of sufficiently flexible stellar density models is crucial for dynamical studies, enabling more accurate predictions of stellar velocity dispersions in a given gravitational potential (Figures \ref{fig:special_beta} and \ref{fig:special_gamma}).  Model selection also impacts inference of stellar orbital energy distributions, which determine resilience to tidal disruption.  

\end{itemize}

One must keep in mind that even our $\alpha\beta\gamma$ model characterizes the dSph density field simplistically, neglecting commonly-observed features like population gradients,  multiple stellar populations, localized substructure, departures from elliptical symmetry \citep[e.g.][]{harbeck01,tolstoy04,coleman05b,okamoto12,wang19,pace20,prabhu26}.  Hercules and Ursa Major II, the two dSphs presenting the most significant stellar cusps, both display the elongated and irregular morphologies that can signify ongoing tidal disruption \citep{munoz10,coleman07b,sand09,munoz18}.  Alternatively, internal heating due to gravitational encounters with subgalactic dark matter halos can cause transient and irregular morphology in the stellar component \citep{penarrubia25}.  \citet{roderick16} identify elongated stellar substructure in the outskirts of \boo\ I, where we measure the shallowest slope of any dSph density profile.  Among the most luminous dSphs, Sculptor, Fornax, Sextans and Ursa Minor all are known to harbor chemodynamically distinct stellar sub-populations, with different spatial configurations.  The extent to which the shape parameters ($\beta,\gamma$) we infer might emerge from such unmodeled complexity is unclear.  Allowing more flexible density profiles at the sub-population level would require simultaneous analysis of spectroscopic data, which is beyond scope of our present work.  

In any case, the composite populations of the Milky Way's dSphs exhibit significant structural diversity that can inform our understanding of their formation and evolution.  Their outer stellar density profiles, characterized here by the log-slope parameter $\beta$, encode the current status of a contest between internal and external gravitational fields.  The steepest slopes that we measure belong to three of the most massive dSphs (Fornax, Leo I and Leo II)---all of which are currently $\gtrsim150$ kpc from the Galactic center---and Eridanus II, which at $D\approx 370$ kpc \citep{pace24} is the most distant dSph in our sample.  This circumstance suggests that steep outer slopes may signify dominance by the internal gravitational field and resilience to tidal perturbation.  Indeed, N-body experiments by \citet{penarrubia09} demonstrate that the outer regions of dSph-like satellites evolve toward shallow power laws after losing stellar mass to external tidal fields.  The steep outer profiles in Fornax, Leo I, Leo II and Eridanus II may then resemble primordial configurations of the dSph stellar component.  A possible exception is Crater II, for which we also infer a steep outer profile (\betaCraterII) and which shows structural and kinematic evidence for ongoing tidal stripping \citep{coppi24,limberg25,atzberger26}, but in that case the posterior PDF for $\beta$ is relatively broad and the evidence ratio (\dlogevidenceCraterII) does not give strong reason to favor $\alpha\beta\gamma$ over the standard Plummer model.

The inner profiles, characterized by log-slope parameter $\gamma$, are likely set by internal processes.  All of the dSphs more massive than $M_{\rm mem}^{\rm tot}\sim 10^5M_{\odot}$ tend to have shallow inner slopes that are statistically compatible with uniform-density cores ($\gamma\approx 0$).  At lower masses, the only well-constrained inner slopes are steep, belonging to Hercules (\gammaHercules) and Ursa Major II (\gammaUrsaMajorII) and, with less compelling evidence in favor of the $\alpha\beta\gamma$ model, \boo\ IV (\gammaBootesIV).  

The cusp we detect in Ursa Major II (UMaII) warrants scrutiny, given previous reports of a faint, central star cluster in that galaxy \citep[][Z06 and E22 hereafter]{zucker06b,eadie22}.  With their original discovery of UMaII, using data from the Sloan Digital Sky Survey, Z06 noted the appearance in followup Subaru/Suprime-Cam imaging of a `bright central concentration' that, like UmaII itself, exhibits a broad main sequence suggesting multiple stellar populations.  Analyzing archival \textit{Hubble Space Telescope} images, E22 classify the central overdensity as a globular cluster, based on its scale radius ($\approx 3$ pc), similarity of its main sequence track to that of UMaII, and a depletion of stars on the lower main sequence---consistent with collisional relaxation/evaporation.  E22 estimate a cluster mass of just $\sim 60 
M_{\odot}$, and note, echoing Z06's original report, that its apparent age/metallicity distribution is similar to that of UMaII.

Even with the overdensity, UMaII's central region is so faint that, while it is listed as a star cluster in the LVDB, by the procedure described in Section \ref{sec:spatialselection} it is left unmasked in our analysis.\footnote{The same is true of the faint star cluster near the center of Eridanus II \citep{koposov15,simon21}, which is unmasked in our analysis.  In that case, though, our estimate of \gammaEridanusII\ is consistent with a core of uniform density in Eridanus II.}  If we loosen the stated surface brightness threshold and mask the four pixels, containing 32 stars, within 30 arcsec of the cluster center (which is separated from UMaII's center by just $\approx 4$ arcsec, as listed in the LVDB), then our fit to UMaII gives $\gamma=0.86\pm 0.40$, unsurprisingly weakening the cusp signal.

So is UMaII's central overdensity a bona fide globular cluster, perhaps in the late stages of tidal disruption, or is it the stellar cusp implied by the fit of our $\alpha\beta\gamma$ model?  Our model, as implemented here, is not so flexible as to account explicitly for a central cluster that is distinct from the main dSph population.  On the other hand, a steep stellar cusp would appear like what is observed in UMaII: a central stellar overdensity that follows the same age/metallicity distribution as the rest of the galaxy and---if it forms via internal dynamical processes \citep[][see below]{errani26}---is depleted of low-mass stars.  Settling the nature of UMaII's central overdensity will require further modeling and perhaps deep spectroscopy to distinguish the stellar populations chemodynamically.  

For the time being, we note that there are several known mechanisms for generating steep stellar cusps dynamically: collisional relaxation of stars in the vicinity of a central black hole \citep{bahcall76}, adiabatic contraction around a slowly growing black hole \citep{gondolo99}, core collapse of halos made from strongly self-interacting dark matter \citep{balberg02}, and/or combinations of these scenarios \citep{shapiro14}.  The longstanding argument against stellar-collisional relaxation in galaxies points to the long timescale for two-body encounters \citep{bt08}; however, recent simulations by \citet{errani26} demonstrate that in the smallest and least massive ultrafaint dSphs, collisional relaxation is accelerated by dynamical friction from sub-solar-mass dark matter particles, prompting mass segregation and formation of baryon-dominated central star clusters.  In order to disentangle the various mechanisms and better understand how steep stellar cusps might form in ultrafaint dSphs, we require simulations that include central black holes and dark matter self-interactions while resolving stellar collisions and dynamical friction from dark matter particles.  We leave this for future work.  

Finally, the detection of steep cusps in ultrafaint dSphs has implications for these galaxies' survival in the external gravitational field of the Milky Way.  Stars comprising cusps are the most strongly bound within a host dark matter halo, populating the lowest energy states (Figure \ref{fig:special_gamma}; see also \citealt{errani24}).  This makes them extraordinarily resilient against tidal stripping, which acts preferentially on weakly bound stars and never fully disrupts cuspy dark matter halos \citep{errani20} or stellar cusps within those halos \citep{errani24}.  Thus dSphs like Hercules and Ursa Major II are plausible progenitors of the `micro galaxies' described by \citet{errani20}.  
 
\section*{Acknowledgments}
M.G.W. and R.E. acknowledge support by National Science Foundation grant AST-2206046.  
E. V. acknowledges funding from the Royal Society, under the Newton International Fellowship programme (NIF\textbackslash R1\textbackslash 241973). S.L.S. acknowledges support from the National
Science Foundation Graduate Research Fellowship Program
under Grant No. DGE2140739.  This material is based upon work supported by the National Aeronautics and Space Administration under Grant/Agreement No. 80NSSC24K0084 as part of the Roman Large Wide Field Science program funded through ROSES call NNH22ZDA001N-ROMAN.

The authors used Claude \citep{Anthropic2026Claude} and ChatGPT \citep[GPT-5.6 Sol;][]{openai2026chatgpt} as tools for copyediting, formatting multi-panel figures and assisting with discussion and refinement of statistical methodology.  The authors generated all text, plots and references manually. 

Funding for SDSS-III has been provided by the Alfred P. Sloan Foundation, the Participating Institutions, the National Science Foundation, and the U.S. Department of Energy Office of Science. The SDSS-III web site is http://www.sdss3.org/.

SDSS-III is managed by the Astrophysical Research Consortium for the Participating Institutions of the SDSS-III Collaboration including the University of Arizona, the Brazilian Participation Group, Brookhaven National Laboratory, Carnegie Mellon University, University of Florida, the French Participation Group, the German Participation Group, Harvard University, the Instituto de Astrofisica de Canarias, the Michigan State/Notre Dame/JINA Participation Group, Johns Hopkins University, Lawrence Berkeley National Laboratory, Max Planck Institute for Astrophysics, Max Planck Institute for Extraterrestrial Physics, New Mexico State University, New York University, Ohio State University, Pennsylvania State University, University of Portsmouth, Princeton University, the Spanish Participation Group, University of Tokyo, University of Utah, Vanderbilt University, University of Virginia, University of Washington, and Yale University.

The Legacy Surveys consist of three individual and complementary projects: the Dark Energy Camera Legacy Survey (DECaLS; Proposal ID 2014B-0404; PIs: David Schlegel and Arjun Dey), the Beijing-Arizona Sky Survey (BASS; NOAO Prop. ID 2015A-0801; PIs: Zhou Xu and Xiaohui Fan), and the Mayall z-band Legacy Survey (MzLS; Prop. ID 2016A-0453; PI: Arjun Dey). DECaLS, BASS and MzLS together include data obtained, respectively, at the Blanco telescope, Cerro Tololo Inter-American Observatory, NSF’s NOIRLab; the Bok telescope, Steward Observatory, University of Arizona; and the Mayall telescope, Kitt Peak National Observatory, NOIRLab. Pipeline processing and analyses of the data were supported by NOIRLab and the Lawrence Berkeley National Laboratory (LBNL). The Legacy Surveys project is honored to be permitted to conduct astronomical research on Iolkam Du’ag (Kitt Peak), a mountain with particular significance to the Tohono O’odham Nation.

NOIRLab is operated by the Association of Universities for Research in Astronomy (AURA) under a cooperative agreement with the National Science Foundation. LBNL is managed by the Regents of the University of California under contract to the U.S. Department of Energy.

This project used data obtained with the Dark Energy Camera (DECam), which was constructed by the Dark Energy Survey (DES) collaboration. Funding for the DES Projects has been provided by the U.S. Department of Energy, the U.S. National Science Foundation, the Ministry of Science and Education of Spain, the Science and Technology Facilities Council of the United Kingdom, the Higher Education Funding Council for England, the National Center for Supercomputing Applications at the University of Illinois at Urbana-Champaign, the Kavli Institute of Cosmological Physics at the University of Chicago, Center for Cosmology and Astro-Particle Physics at the Ohio State University, the Mitchell Institute for Fundamental Physics and Astronomy at Texas A\&M University, Financiadora de Estudos e Projetos, Fundacao Carlos Chagas Filho de Amparo, Financiadora de Estudos e Projetos, Fundacao Carlos Chagas Filho de Amparo a Pesquisa do Estado do Rio de Janeiro, Conselho Nacional de Desenvolvimento Cientifico e Tecnologico and the Ministerio da Ciencia, Tecnologia e Inovacao, the Deutsche Forschungsgemeinschaft and the Collaborating Institutions in the Dark Energy Survey. The Collaborating Institutions are Argonne National Laboratory, the University of California at Santa Cruz, the University of Cambridge, Centro de Investigaciones Energeticas, Medioambientales y Tecnologicas-Madrid, the University of Chicago, University College London, the DES-Brazil Consortium, the University of Edinburgh, the Eidgenossische Technische Hochschule (ETH) Zurich, Fermi National Accelerator Laboratory, the University of Illinois at Urbana-Champaign, the Institut de Ciencies de l’Espai (IEEC/CSIC), the Institut de Fisica d’Altes Energies, Lawrence Berkeley National Laboratory, the Ludwig Maximilians Universitat Munchen and the associated Excellence Cluster Universe, the University of Michigan, NSF’s NOIRLab, the University of Nottingham, the Ohio State University, the University of Pennsylvania, the University of Portsmouth, SLAC National Accelerator Laboratory, Stanford University, the University of Sussex, and Texas A\&M University.

BASS is a key project of the Telescope Access Program (TAP), which has been funded by the National Astronomical Observatories of China, the Chinese Academy of Sciences (the Strategic Priority Research Program “The Emergence of Cosmological Structures” Grant XDB09000000), and the Special Fund for Astronomy from the Ministry of Finance. The BASS is also supported by the External Cooperation Program of Chinese Academy of Sciences (Grant 114A11KYSB20160057), and Chinese National Natural Science Foundation (Grant 12120101003, 11433005).

The Legacy Survey team makes use of data products from the Near-Earth Object Wide-field Infrared Survey Explorer (NEOWISE), which is a project of the Jet Propulsion Laboratory/California Institute of Technology. NEOWISE is funded by the National Aeronautics and Space Administration.

The Legacy Surveys imaging of the DESI footprint is supported by the Director, Office of Science, Office of High Energy Physics of the U.S. Department of Energy under Contract No. DE-AC02-05CH1123, by the National Energy Research Scientific Computing Center, a DOE Office of Science User Facility under the same contract; and by the U.S. National Science Foundation, Division of Astronomical Sciences under Contract No. AST-0950945 to NOAO.

This project used public archival data from the Dark Energy Survey (DES). Funding for the DES Projects has been provided by the U.S. Department of Energy, the U.S. National Science Foundation, the Ministry of Science and Education of Spain, the Science and Technology FacilitiesCouncil of the United Kingdom, the Higher Education Funding Council for England, the National Center for Supercomputing Applications at the University of Illinois at Urbana-Champaign, the Kavli Institute of Cosmological Physics at the University of Chicago, the Center for Cosmology and Astro-Particle Physics at the Ohio State University, the Mitchell Institute for Fundamental Physics and Astronomy at Texas A\&M University, Financiadora de Estudos e Projetos, Funda{\c c}{\~a}o Carlos Chagas Filho de Amparo {\`a} Pesquisa do Estado do Rio de Janeiro, Conselho Nacional de Desenvolvimento Cient{\'i}fico e Tecnol{\'o}gico and the Minist{\'e}rio da Ci{\^e}ncia, Tecnologia e Inova{\c c}{\~a}o, the Deutsche Forschungsgemeinschaft, and the Collaborating Institutions in the Dark Energy Survey.
The Collaborating Institutions are Argonne National Laboratory, the University of California at Santa Cruz, the University of Cambridge, Centro de Investigaciones Energ{\'e}ticas, Medioambientales y Tecnol{\'o}gicas-Madrid, the University of Chicago, University College London, the DES-Brazil Consortium, the University of Edinburgh, the Eidgen{\"o}ssische Technische Hochschule (ETH) Z{\"u}rich,  Fermi National Accelerator Laboratory, the University of Illinois at Urbana-Champaign, the Institut de Ci{\`e}ncies de l'Espai (IEEC/CSIC), the Institut de F{\'i}sica d'Altes Energies, Lawrence Berkeley National Laboratory, the Ludwig-Maximilians Universit{\"a}t M{\"u}nchen and the associated Excellence Cluster Universe, the University of Michigan, the National Optical Astronomy Observatory, the University of Nottingham, The Ohio State University, the OzDES Membership Consortium, the University of Pennsylvania, the University of Portsmouth, SLAC National Accelerator Laboratory, Stanford University, the University of Sussex, and Texas A\&M University.
Based in part on observations at Cerro Tololo Inter-American Observatory, National Optical Astronomy Observatory, which is operated by the Association of Universities for Research in Astronomy (AURA) under a cooperative agreement with the National Science Foundation.

This work has made use of data from the European Space Agency (ESA) mission
{\it Gaia} (\url{https://www.cosmos.esa.int/gaia}), processed by the {\it Gaia}
Data Processing and Analysis Consortium (DPAC,
\url{https://www.cosmos.esa.int/web/gaia/dpac/consortium}). Funding for the DPAC
has been provided by national institutions, in particular the institutions
participating in the {\it Gaia} Multilateral Agreement.

The Pan-STARRS1 Surveys (PS1) and the PS1 public science archive have been made possible through contributions by the Institute for Astronomy, the University of Hawaii, the Pan-STARRS Project Office, the Max-Planck Society and its participating institutes, the Max Planck Institute for Astronomy, Heidelberg and the Max Planck Institute for Extraterrestrial Physics, Garching, The Johns Hopkins University, Durham University, the University of Edinburgh, the Queen's University Belfast, the Harvard-Smithsonian Center for Astrophysics, the Las Cumbres Observatory Global Telescope Network Incorporated, the National Central University of Taiwan, the Space Telescope Science Institute, the National Aeronautics and Space Administration under Grant No. NNX08AR22G issued through the Planetary Science Division of the NASA Science Mission Directorate, the National Science Foundation Grant No. AST-1238877, the University of Maryland, Eotvos Lorand University (ELTE), the Los Alamos National Laboratory, and the Gordon and Betty Moore Foundation.

\appendix

Data products associated with this work are publicly available at the Zenodo database, https://doi.org/10.5281/zenodo.20600393.  For every galaxy/survey pair, we provide an input catalog file named `[galaxy name]\_[survey name]\_catalog.fits', and output files named `[galaxy name]\_[survey name]\_[model]\_posterior.fits', with separate output files for models identified as `abg' ($\alpha\beta\gamma$), `plum' (Plummer), `\sersic' and `exp' (exponential).  Output files corresponding to the spherical ($\epsilon=0$) versions of these models have names `[galaxy name]\_[survey name]\_[model]\_sph\_posterior.fits'.  Tables \ref{tab:input} and \ref{tab:output} list the contents of the input and output files, which all have multi-extension FITS format.  Some information (e.g., the LVDB record for the specified galaxy) is included in both the input and output files, accommodating those who may be interested in only one or the other type.  

The files named `[galaxy name]\_[survey name]\_[model]\_sph\_posterior.fits' contain random samples from posterior PDFs.   We compute the 3D halflight radius, $r_{\rm half}$, as that which satisfies $4\pi\int_0^{r_{\rm half}}r^2\nu_{\rm mem}(r)dr=\frac{1}{2}N_{\rm mem}^{\rm obs}(\infty)$, which is strictly valid only for the spherical models.  Because it has been used in dynamical mass estimators \citep{wolf10}, we also compute $r_{-3}$, the 3D radius at which the stellar density profile has slope $d\log{\nu}/d\log r=-3$.  In order to list these quantities, as well as the circularized halflight radius, $R_{\rm half}=a_{\rm half}\sqrt{1-\epsilon}$, in units of pc, we sample dSph distance moduli from a Gaussian distribution with mean and standard deviation as listed in the LVDB.

The single file named `table.fits' is an expanded version of Table \ref{tab:sbfit_table}, in binary FITS format, listing summary statistics (mean, standard deviation, 16th and 84th percentiles) of posterior PDFs for all model and implied parameters for all fits to all catalogs.  

Finally, the Zenodo database includes figures (PDF format) similar that we have displayed here for \boo\ I, but for all dSphs.  Table \ref{tab:figures} lists the different kinds of figures by filename and explains what they depict.

\begin{deluxetable}{lll}
\tablewidth{0pt}
\tablecaption{Contents of electronic input source catalogs `[galaxy name]\_[survey name]\_catalog.fits'
\label{tab:input}
}
\tablehead{\colhead{\shortstack{[Extension].Header/\\Data Unit}}&\colhead{keyword/column\tablenotemark{*}}&\colhead{description}}
\startdata
$[0]$.header & object & galaxy name\\
$[0]$.header & survey & survey name\\
$[0]$.header & grid\_dimension & dimensions of pixel grid array\\
$[0]$.header & pixel\_length/arcmin & angular side length of one pixel in grid array\\
$[0]$.header & R\_min/Rhalf\_published & minimum separation from nominal center (to avoid crowding)\\
$[0]$.header & log10(age/years) & age adopted for isochrone\\
$[0]$.header & Fe/H & metallicity adopted for isochrone\\
$[0]$.header & isochrone\_color\_offset & offset added to isochrone color\\
$[0]$.header & mag\_limit\_bright & bright-end magnitude limit\\
$[0]$.header & mag\_limit\_faint & faint-end magnitude limit\\
$[0]$.header & n\_sources & number of unmasked sources included in fit\\
$[0]$.header & isochrone\_color & filters used for isochrone color\\
$[0]$.header & isochrone\_magnitude & filter used for isochrone magnitude\\
$[1]$.data &\nodata & LVDB record for [galaxy\_name], columns defined by \citet{pace24}\\
$[2]$.data & ra & right ascension in astrometric system of survey catalog [deg]\\
$[2]$.data & dec & declination in astrometric system of survey catalog [deg]\\
$[2]$.data & x & $\xi$ coordinate in tangent plane defined by LVDB centroid [deg]\\
$[2]$.data & y & $\eta$ coordinate in tangent plane defined by LVDB centroid [deg]\\
$[2]$.data & $X$mag & magnitude in $X$ filter, uncorrected for extinction  ($X$ can be $ugrizy$ and $G,BP,RP$)\\
$[2]$.data & e\_$X$mag & error in magnitude in $X$ filter\\
$[2]$.data & ext\_$X$mag & extinction correction to be applied to $X$mag, corrected magnitude is \texttt{$X$\_mag$-$ext\_$X$mag}\\
$[2]$.data & cmd\_mask & True for stars outside color/magnitude selection\\
$[2]$.data & pos\_mask & True for stars with position outside field or within masked pixels\\
$[3]$.data &\nodata & 2D image array containing right ascension coordinate at center of pixel [deg]\\
$[4]$.data &\nodata & 2D image array containing declination coordinate at center of pixel [deg]\\
$[5]$.data &\nodata & 2D image array containing value of pos\_mask at each pixel\\
$[6]$.data & isochrone\_mag & (apparent) magnitude array for adopted isochrone\\
$[6]$.data & isochrone\_col & color array for adopted isochrone\\
\enddata
\tablenotetext{*}{Entries with `.header' use FITS header keywords.  Entries with `[2].data' or `[6].data' use binary table column names.  Entries with `[3].data', `[4].data', `[5].data' are 2D image arrays.}
\end{deluxetable}

\clearpage
\startlongtable
\begin{deluxetable*}{lll}
\tablewidth{0pt}
\tablecaption{Contents of electronic output files `[galaxy\_name]\_[survey\_name]\_[model]\_posterior.fits'
\label{tab:output}
}
\tablehead{\colhead{\shortstack{[Extension].Header/\\Data Unit}}&\colhead{keyword/column\tablenotemark{*}}&\colhead{description}}
\startdata
$[0]$.header & object & galaxy name\\
$[0]$.header & survey & survey name\\
$[0]$.header & model & adopted model (abg/plum/sersic/exp, with \_sph indicating spherical version)\\
$[0]$.header & grid\_dimension & dimensions of pixel grid array\\
$[0]$.header & pixel\_length/arcmin & angular side length of one pixel in grid array\\
$[0]$.header & R\_min/Rhalf\_published & minimum separation from nominal center (to avoid crowding)\\
$[0]$.header & log10(age/years) & age adopted for isochrone\\
$[0]$.header & Fe/H & metallicity adopted for isochrone\\
$[0]$.header & isochrone\_color\_offset & offset added to isochrone color\\
$[0]$.header & mag\_limit\_bright & bright-end magnitude limit\\
$[0]$.header & mag\_limit\_faint & faint-end magnitude limit\\
$[0]$.header & n\_sources & number of unmasked sources included in fit\\
$[0]$.header & isochrone\_color & filters used for isochrone color\\
$[0]$.header & isochrone\_magnitude & filter used for isochrone magnitude\\
$[0]$.header & ln(evidence) & log-evidence, returned by MultiNest\\
$[0]$.header & e\_ln(evidence) & error in log-evidence, returned by MultiNest\\
$[0]$.header & sn & signal-to-noise ratio of galaxy overdensity\\
$[0]$.header & chi2dof & $\chi^2$ per degree of freedom, estimated from binned profile\\
$[0]$.header & bkd\_obs & empirical estimate of background stellar density [stars arcmin$^{-2}$]\\
$[0]$.header & bkd\_bestfit & best-fitting $\Sigma_{\rm non,0}$ [stars arcmin$^{-2}$]\\
$[1]$.data &\nodata & LVDB record for [galaxy\_name], columns defined by \citet{pace24}\\
$[2]$.data & dx & $\xi_0$, offset of centroid from LVDB-listed value in nominal tangent plane [deg]\\
$[2]$.data & dy & $\eta_0$, offset of centroid from LVDB-listed value in nominal tangent plane [deg]\\
$[2]$.data & n\_in\_field & $N_{\rm field}$, (model-predicted) number of stars in the observed field\\
$[2]$.data & member\_fraction\_in\_field & $f_{\rm mem}$, member fraction among observed stars\\
$[2]$.data & r\_scale & $r_s$, scale radius [arcmin]\\
$[2]$.data & ellipticity & $\epsilon$, ellipticity\\
$[2]$.data & position\_angle & $\theta$, position angle [deg]\\
$[2]$.data & alpha & $\alpha$, sharpness of power-law break\\
$[2]$.data & beta & $\beta$, outer power-law index\\
$[2]$.data & gamma & $\gamma$, inner power-law index\\
$[2]$.data & bkd\_grad\_x & $a_{\xi}$, $\xi$-component of foreground density gradient [$R_{h,pub}^{-1}$]\\
$[2]$.data & bkd\_grad\_y & $a_{\eta}$, $\eta$-component of foreground density gradient [$R_{h,pub}^{-1}$]\\
$[2]$.data & ra & $\alpha_0$, centroid right ascension, in astrometric system of survey catalog [deg]\\
$[2]$.data & dec & $\delta_0$, centroid declination, in astrometric system of survey catalog [deg]\\
$[2]$.data & Sigma0\_mem & $\Sigma_0$, normalization of member surface density profile [stars arcmin$^{-2}$]\\
$[2]$.data & Sigma0\_non & $\Sigma_{\rm non,0}$, surface density scale of foreground/background nonmembers [stars arcmin$^{-2}$]\\
$[2]$.data & nu\_scale & $\nu_s$, normalization of 3D density profile [stars arcmin$^{-3}$]\tablenotemark{**}\\
$[2]$.data & a\_half & $a_{\rm half}$, halflight elliptical radius [arcmin] \\
$[2]$.data & n\_member\_observed & $N_{\rm mem}^{\rm obs}(\infty)$, number of members within adopted magnitude limits\\
$[2]$.data & n\_member\_observed\_in\_field & $f_{\rm mem}N_{\rm field}$, number of members within adopted magnitude and field limits\\
$[2]$.data & distance\_modulus & distance modulus (sampled from LVDB-listed interval) for converting 2D to 3D radii\\
$[2]$.data & r\_half\_2d & $R_{\rm half}$, circularized (2D) halflight radius [pc]\\
$[2]$.data & r\_half\_3d & $r_{\rm half}$, (3D) halflight radius [pc]\\
$[2]$.data & r\_3 & $r_3$, (3D) radius where $d\log\nu/d\log r=-3$ [pc]\\
$[2]$.data & loglikelihood & $\ln P(D|\vec{\theta})$, natural logarithm of the data likelihood, given model $\vec{\theta}$\\
$[2]$.data & kroupa\_n\_tot & $N_{\rm mem}^{\rm tot}$, total number of member stars, Kroupa mass function\\
$[2]$.data & kroupa\_m\_tot & $M_{\rm mem}^{\rm tot}$, total stellar mass, Kroupa mass function [$M_{\odot}$]\\
$[2]$.data & kroupa\_l\_tot & $L_{\rm mem}^{\rm tot}$, total stellar luminosity, Kroupa mass function [$L_{\odot}$]\\
$[2]$.data & kroupa\_lv\_tot & $L_{\rm V,mem}^{\rm tot}$, total stellar V-band luminosity, Kroupa mass function [$L_{\odot}$]\\
$[2]$.data & kroupa\_M\_V & $M_V$, galaxy absolute magnitude, Kroupa mass function\\ 
$[2]$.data & salpeter\_n\_tot & $N_{\rm mem}^{\rm tot}$, total number of member stars, Salpeter mass function\\
$[2]$.data & salpeter\_m\_tot & $M_{\rm mem}^{\rm tot}$, total stellar mass, Salpeter mass function [$M_{\odot}$]\\
$[2]$.data & salpeter\_l\_tot & $L_{\rm mem}^{\rm tot}$,  total stellar luminosity, Salpeter mass function [$L_{\odot}$]\\
$[2]$.data & salpeter\_lv\_tot & $L_{\rm V,mem}^{\rm tot}$, total stellar V-band luminosity, Salpeter mass function [$L_{\odot}$]\\
$[2]$.data & salpeter\_M\_V & $M_V$, galaxy absolute magnitude, Salpeter mass function\\ 
\enddata
\tablenotetext{*}{Entries with `.header' use FITS header keywords.  Entries with `[2].data' use binary table column names.}
\tablenotetext{**}{The authors acknowledge the absurdity of this unit, which is necessary for the projection of $\nu(r)$ to have the standard unit [stars arcmin$^{-2}$].}
\end{deluxetable*}

\clearpage
\begin{deluxetable}{lll}
\tablewidth{0pt}
\tablecaption{Contents of figure files in Zenodo database
\label{tab:figures}
}
\tablehead{\colhead{File name}&\colhead{description}}
\startdata
[galaxy\_name]\_figure\_cmd.pdf&color-magnitude diagrams for all surveys, see Figure \ref{fig:boo1_figure}, top row\\
$[$galaxy\_name]\_figure\_sb\_obs\_grid.pdf&observed stellar density field, see Figure \ref{fig:boo1_figure}, second row\\
$[$galaxy\_name]\_figure\_mask\_grid.pdf&pixel mask, see Figure \ref{fig:boo1_figure}, third row\\
$[$galaxy\_name]\_figure\_sb\_bestfit\_grid\_abg.pdf&best-fitting $\alpha\beta\gamma$ model stellar density field, see Figure \ref{fig:boo1_figure}, fourth row\\
$[$galaxy\_name]\_figure\_sb\_residual\_grid\_abg.pdf&normalized residuals with respect to best-fitting $\alpha\beta\gamma$ model, see Figure \ref{fig:boo1_figure}, fifth row\\
$[$galaxy\_name]\_figure\_sb\_profile\_abg.pdf&binned stellar density profile with $\alpha\beta\gamma$ posterior overplotted, see Figure \ref{fig:boo1_figure}, sixth row\\
$[$galaxy\_name]\_[survey\_name]\_sb\_corner1\_abg.pdf & corner plot for $\alpha\beta\gamma$ model parameters, see Figure \ref{fig:Bootes_I_decals_sb_corner1_abg}\\
$[$galaxy\_name]\_[survey\_name]\_sb\_corner2\_abg.pdf & corner plot for ($\alpha\beta\gamma$ model) implied parameters, see Figure \ref{fig:Bootes_I_decals_sb_corner2_abg}\\
$[$galaxy\_name]\_profiles.pdf&stellar density, velocity dispersion profiles and energy distribution, see Figures \ref{fig:special_beta}-\ref{fig:special_gamma}\\
\enddata
\end{deluxetable}

\bibliography{ref}
\end{document}